\documentstyle[aps,prl]{revtex}
\input epsf.sty
\begin{document}

\twocolumn[\hsize\textwidth\columnwidth\hsize\csname@twocolumnfalse\endcsname

\draft
\title{Random-Mass Dirac Fermions in an Imaginary 
Vector Potential (I): \\
Delocalization Transition and Localization Length}
\author{Koujin Takeda\cite{Takeda}}
\address{Institute for Cosmic Ray Research, University of Tokyo, Kashiwa, 
Chiba, 277-8582 Japan
}
\author{Ikuo Ichinose\cite{Ichinose}}
\address{Department of Electrical and Computer Engineering, 
Nagoya Institute of Technology, Gokiso, Showa-ku, 
Nagoya, 466-8555 Japan
}
%\date{, 2000}
\maketitle

\begin{abstract}
In this and subsequent papers,
one dimensional system of Dirac fermions with a random-varying mass 
is studied by the transfer-matrix methods which we developed recently.
We investigate the  effects of nonlocal correlation of the spatial-varying 
Dirac mass on the delocalization transition.
Especially we numerically calculate both the ``typical" and 
``mean" localization
lengths as a function of energy and the correlation length of 
the random mass.
To this end we introduce an imaginary vector potential as suggested
by Hatano and Nelson and solve
the eigenvalue problem.
Numerical calculations are in good agreement with the results 
of the analytical calculations.
We obtain the relation between the localization length of states
and the correlation length of the random mass.
In subsequent paper, we shall study Dirac fermions with a random mass
of long-ranged correlations.
\end{abstract}


]

%\begin{multicols}{2}
\section{Introduction}
One of the most important problems in condensed matter
physics is the localization phenomena in random-disordered 
systems\cite{Anderson}.
At present it is believed that almost all states tend to localize
in disorder systems in two and lower dimensions. 
In some special cases, however, some specific states remain
extended even in the presence of strong disorders.  
System of Dirac fermions with a random-varying mass(RMDF) in one
dimension has been studied from this point of 
view\cite{Ovc,Comtet,DSF,BF,Mathur,Steiner,Shelton,Gog}.
RMDF is a continuum field theory of the random hopping tight
binding model and it is related with the random XY spin chain,
the random Ising model, polyacetylene system, etc.

In most of studies on the random-disordered systems, only the 
short-range white-noise random variables are considered.
In the previous papers\cite{TTIK,IK,IK2} we studied the 
effect of nonlocal correlation
of the random mass on the extended states which exist near the band
center.
For numerical studies, we reformulate the system by transfer-matrix 
formalism, and obtained
eigenvalues and wave functions for various configurations
of random telegraphic mass\cite{TTIK}.
We verified that the density of states obtained by the transfer-matrix 
methods is in good agreement with the analytical calculation 
by supersymmetric methods in Ref.\cite{IK}.
In this paper we shall introduce an imaginary vector potential into
the system of the RMDF and study a 
localization-delocalization phase transition by varying the magnitude of
the vector potential.
Through this study we shall obtain the localization 
length of the states quite accurately as 
a function of energy and the correlation length of the random mass.
This method of calculating the localization length is based on the idea
by Hatano and Nelson\cite{HN} and it is totally different from 
the methods which have been used so far\cite{Mac}.

In this and subsequent papers we shall study effects of 
long-range spatial correlations of the random mass, especially on
delocalized states, localization lengths, etc.
This work is partially motivated by the recent studies on the 
one-dimensional Anderson model with random potentials with long-range
correlation.
There it is shown that there exists a mobility edge at a finite
energy if the correlation is long-range enough.
As we explained above, extended states exist at the band center
in the present model even for the white-noise random mass.
Then it is interesting to see if additional extended states appear as a 
result of the long-range correlation of the random mass.

This paper is organized as follows.
In Sec.2, we first explain the imaginary-vector potential
methods(IVPM) for calculating the typical and mean localization lengths.
Then we compare numerical results with the analytic calculations
in order to verify the validity of the IVPM.
In Sec.3, we shall consider the system of the RMDF in
which a Gaussian distribution is employed
for distances between (anti-)kinks or jumps of the telegraphic
random mass.
We calculate the correlation length of the random mass
and also the localization lengths as a function of the mean distance
and standard deviation.
Then we obtain relation between the correlation length of the 
random mass and the localization lengths.
Section 4 is devoted to conclusion.

The studies on the Anderson model etc., in one dimension employ
specific techniques like the Hamiltonian mapping, the renormalization
group, etc.
Compared with them, our method are rather straightforward.
Before calculating the localization lengths, we obtain landscapes
of the wave functions which are useful to get physical pictures of the
delocalization transition.

%%%%%%%%%%%%%%%%%%%%%%%%%%%%%%%%%%%%%%%%%%%%%%%%%%%%%%%%%%%%%%%%%%%%

\section{Model, TMM and IVPM} 
We shall consider a Dirac fermion in one spatial dimension with a 
coordinate-dependent mass $m(x)$ and in an imaginary vector potential $g$,
whose Hamiltonian is given by,
\begin{eqnarray}
{\cal H}&=&\int dx \psi^\dagger h\psi,\\
h&=&-i\sigma^z (\partial_x+g) +m(x)\sigma^y,
\end{eqnarray}
where $\vec{\sigma}$ are the Pauli matrices. 
This fermion model is a low-energy effective model of random-hopping 
tight binding models, random-bond spin chains, etc\cite{BF}.
We introduce the components of $\psi$ as $\psi=(u,v)$.
In terms of them the Dirac equation is given as,
\begin{eqnarray}
\left(\ \frac{d}{dx}\ +\ g +\ m(x)\ \right) u(x)&=&Ev(x),\nonumber\\
\left(\ -\frac{d}{dx}\ -\ g +\ m(x)\ \right) v(x)&=&Eu(x). 
\label{eq:dirac1}
\end{eqnarray}
(We follow the notations in Ref.\cite{Comtet}.) 
From Eqs.(\ref{eq:dirac1}), we obtain the
Schr$\ddot{\mbox{o}}$dinger equations,
\begin{eqnarray}
 && \left(-\frac{d^2}{dx^2} - 2g \frac{d}{dx} -m'(x)+
(m^2(x)-g^2)\right) u(x)
  \nonumber \\ 
 &&\hspace{5cm} = E^2u(x),
\label{eq:schroedinger1}
\end{eqnarray}
and similarly for $v(x)$.

In this paper we restrict the shapes of $m(x)$ to
multi-kink-anti-kink configurations\cite{NS}. 
The multi-soliton-antisoliton configurations
are given by,
\begin{eqnarray}
m(x)&=&\sum_i\bar{m}(\theta(x-\alpha_i)-1) \nonumber  \\
&&+\sum_j\bar{m}
(\theta(-x+\beta_j)-1),
\label{stepm}
\end{eqnarray}
where $\alpha_i$'s($\beta_j$'s) are positions of kinks(anti-kinks)
and they are random variables in the present system.
An example of $m(x)$ is given in Fig.1.
\begin{figure}
\begin{center}
\unitlength=1cm

\begin{picture}(15,4)
\centerline{
\epsfysize=4cm
\vspace{0mm}
\epsfbox{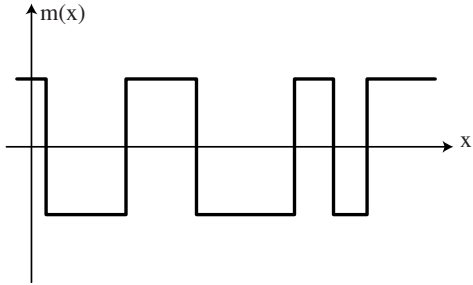}
}
\end{picture}
\vspace{0mm}
\caption{An example of configurations of solitons and anti-solitons.}
\end{center}
\label{fig:mx}
\end{figure}%

If we vary the distances $l$ between soliton and anti-soliton
according to the exponential distribution like,
\begin{equation}
P(l)={1 \over 2\tilde{\lambda}} \exp 
\Big(-{l\over 2\tilde{\lambda}}\Big),
\label{Pl}
\end{equation}
where $\tilde{\lambda}$ is a parameter, then 
$m(x)$ has the following correlation\cite{Comtet},
 \begin{equation}
 [\ m(x)\ m(y)\ ]_{{\rm ens}} = \frac{A}{\tilde{\lambda}} 
 \exp\ (-|x-y| / \tilde{\lambda}),
 \label{disordercor}
 \end{equation} 
where $\sqrt{\frac{A}{\tilde{\lambda}}}$ essentially corresponds to 
the height of the soliton and 
anti-soliton, i.e., $\bar{m}$ in (\ref{stepm}).
From (\ref{disordercor}), $\tilde{\lambda}$ is the correlation length
of the random mass and the limit $\tilde{\lambda} \rightarrow 0$ corresponds
to the white-noise case.
In the subsequent papers\cite{TI,TI2}, we shall study Dirac fermions
with long-range correlated random mass by using the methods
examined in this paper.
There we expect some interesting phenomena like existence
of nontrivial mobility edge, nonuniversality of the multi-fractal
exponents, etc.
Studies in this section show that the IVPM
for calculating the localization lengths are reliable and we shall
use them for studies on random systems with long-range correlated
disorder.

For the vanishing imaginary vector potential,
we can solve the Schr$\ddot{\mbox{o}}$dinger equations
 in (\ref{eq:schroedinger1})
under the periodic boundary condition with various multi-soliton-antisoliton 
configurations of $m(x)$ using the transfer-matrix method(TMM). 
We can also obtain the energy spectrum and wave functions\cite{TTIK} using
 this method.

We review the TMM method briefly.
Let us consider the Dirac equation (\ref{eq:dirac1}) with a mass 
configuration 
which has a kink at $x=0$,
\begin{equation}
m(x)=\bar{m}(2\theta(-x)-1)+m_0,\label{eq:stepfunct}
\end{equation}
where  $\theta(x)$ is the Heaviside function,
\begin{equation}
\theta(x)=\left\{
\begin{array}{ccc}
0&\ & (x < 0) \\
1&\ & (x > 0)
\end{array}
\right. ,
\end{equation}
and it is assumed that
\[\bar{m}>m_0>0.\]
Substituting (\ref{eq:stepfunct}) into (\ref{eq:schroedinger1}), we obtain 
a Schr$\ddot{\mbox{o}}$dinger equation with a potential which is 
a combination of 
the delta function and the step function,
\begin{eqnarray}
V(x) &=& \mp m'(x)+m^2(x)=\pm2\bar{m}\delta(x)+4m_0\bar{m}
\theta(-x) \nonumber \\
 && +(\bar{m}-m_0)^2.
\end{eqnarray}
For such a field equation with a delta-function-type potential, 
the continuity of the wavefunction $u(x)$ leads automatically to 
the conditions 
on its derivative $u'(x)$. 
For example, if we rewrite the Schr$\ddot{\mbox{o}}$dinger equation as
\begin{eqnarray}
\left(-\frac{d^2}{dx^2}+2\bar{m}\ \delta(x)+4m_0\bar{m}\ \theta(-x)\right)
\ u(x) \nonumber \\
 = (E^2-(\bar{m}-m_0)^2)\ u(x), 
\label{Seq}
\end{eqnarray}
(here we restrict $g$ to zero)
and integrate Eq.(\ref{Seq}) with respect to $x$ in the range
$-\epsilon\le x\le\epsilon$, we have
\begin{eqnarray}
&& -u'(0+\epsilon)+u'(0-\epsilon)+2\bar{m}\ u(0) 
+4m_0\bar{m}\int^0_{\epsilon}
 dx\ u(x) \nonumber \\
&& = (E^2-(\bar{m}-m_0)^2)\int^{\epsilon}_{-\epsilon}dx\ u(x). 
\label{conti}
\end{eqnarray}
We see the {\it r.h.s.} of Eq.(\ref{conti}) vanishes in the limit 
$\epsilon\to0$ due to 
the continuity of $u(x)$. 
A similar argument applies to $v(x)$ and we have conditions,
\begin{eqnarray}
u'(0+0)-u'(0-0)&=&2\bar{m}u(0),\nonumber\\
v'(0+0)-v'(0-0)&=&-2\bar{m}v(0).\label{eq:BC1}
\end{eqnarray}
We look for solutions under these conditions. 

Let the energy eigenvalue $E$ satisfy
\begin{equation}
\bar{m}-m_0>E>0,
\end{equation}
then $u(x)$ takes the form,
\begin{eqnarray}
u(x)&=&\left\{
\begin{array}{ccc}
A\ e^{\kappa_1 x}+B\ e^{-\kappa_1 x}&\ &(x<0),\\
C\ e^{\kappa_2 x}+D\ e^{-\kappa_2 x}&\ &(x>0),
\end{array}
\right.\\
\kappa_1&=&\sqrt{(\bar{m}+m_0)^2-E^2},\\
\kappa_2&=&\sqrt{(\bar{m}-m_0)^2-E^2}.
\end{eqnarray}
From the continuity of $u(x)$,
\begin{equation}
A+B=C+D,
\end{equation}
and Eq.(\ref{eq:BC1}) gives
\begin{equation}
\kappa_2(C-D)-\kappa_1(A-B)=2\bar{m}(A+B).
\end{equation}
Hence,
\begin{equation}
u(x)=Ae^{\kappa_1x}+Be^{-\kappa_1x},\hspace{4mm}{\rm for}\hspace{4mm}x<0\\
\end{equation}
and for $x>0$,
\begin{eqnarray}
u(x)=\left[\ \frac{2\bar{m}+\kappa_1+\kappa_2}{2\kappa_2}A+\frac{2\bar{m}
-\kappa_1
+\kappa_2}{2\kappa_2}B\ \right]\ e^{\kappa_2x}\nonumber\\
\nonumber\\
+\left[\ \frac{-2\bar{m}-\kappa_1+\kappa_2}{2\kappa_2}A+\frac{-2\bar{m}
+\kappa_1
+\kappa_2}{2\kappa_2}B\ \right]\ e^{-\kappa_2x}.
\label{eq:solution4}
\end{eqnarray}
If we let $m_0=0$, $\kappa_1=\kappa_2 \equiv \kappa$,
Eq.(\ref{eq:solution4}) simplifies to
\begin{equation}
u(x)=\left\{
\begin{array}{cc}
A\ e^{\kappa x}+B\ e^{-\kappa x}, & (x<0) \\
 \begin{array}{c}
 \displaystyle{\left[\ \frac{\bar{m}+\kappa}{\kappa}A+
 \frac{\bar{m}}{\kappa}B\ \right]\ e^{\kappa x}}  \\
 \hspace{8mm} +\displaystyle{\left[\ -\frac{\bar{m}}{\kappa}A
 +\frac{-\bar{m}+\kappa}{\kappa}B\ \right]e^{-\kappa x}}. 
 \end{array}
 & (x>0)
\end{array}
\right.
\end{equation}
Values of $A,\ B$ and $E$ are determined by boundary conditions, 
{\it e.g.}, $u(-L/2)=u(L/2)=0$\ for a given system size $L$.

We restrict $m_0$ to zero in the rest of the discussion. 
Since the wavefunction  $u(x)$ is expressed everywhere as
\begin{equation}
u(x)=A\ e^{\kappa x}+B\ e^{-\kappa x},
\end{equation}
we can represent the eigenfunction in terms of coefficients $A$ and $B$. 
By using matrix representation and  from (\ref{eq:solution4}), the  
conditions (\ref{eq:BC1}) give the following relation between 
the coefficients $A$ and $B$,
\begin{eqnarray}
\left(
\begin{array}{c}
A(x>0)\\
B(x>0)
\end{array}
\right)&=&T\left(
\begin{array}{c}
A(x<0)\\
B(x<0)
\end{array}
\right), \nonumber\\
&&\nonumber\\
T&=&\left(
\begin{array}{cc}
\displaystyle{1+\frac{\bar{m}}{\kappa}}&\displaystyle{\frac{\bar{m}}{\kappa}}\\
&\\
\displaystyle{-\frac{\bar{m}}{\kappa}}&\displaystyle{1-\frac{\bar{m}}{\kappa}}
\end{array}
\right).\label{eq:defineT}
\end{eqnarray}
For  kink instead of anti-kink, we should replace $\bar{m}$ with 
$(-\bar{m})$ in Eq.(\ref{eq:defineT}) and we obtain
\begin{equation}
\left(
\begin{array}{c}
A(x>0)\\
B(x>0)
\end{array}
\right)=T^{-1}\left(
\begin{array}{c}
A(x<0)\\
B(x<0)
\end{array}
\right).
\end{equation}
Let us define
\begin{eqnarray}
R(\kappa,a)&\equiv&\left(
\begin{array}{cc}
e^{\kappa a}&0\\
0&e^{-\kappa a}
\end{array}
\right),\nonumber\\
\phi&\equiv&\frac{\bar{m}}{\kappa},
\end{eqnarray}
and TM for the configuration of  an anti-kink and a kink is
given by 
\begin{eqnarray}
\lefteqn{R(b)T^{-1}R(a)T}\nonumber\\
&=&R(b)\left(
\begin{array}{cc}
e^{\kappa a}-2\phi^2\sinh\kappa a&2\phi(1-\phi)\sinh\kappa a\\
2\phi(1+\phi)\sinh\kappa a&e^{-\kappa a}+2\phi^2\sinh\kappa a
\end{array}
\right)\nonumber\\
&\equiv&T(a,b),\label{eq:tmatrtix1}
\end{eqnarray}
where $a$ is the distance between the kink and anti-kink.

We impose {\it periodic} boundary condition on the wavefunction.
The above argument can be generalized to an arbitrary number of 
kink and anti-kink 
pairs readily and we have the following equation
expressed by the transfer matrices.
\begin{equation}
\label{eq:tmatrix}
\mbox{\Large det} \left[ T(e,f)\cdots T(c,d)T(a,b) - \left(
\begin{array}{cc}
1&0 \\
0&1 \\
\end{array}
\right) \right] = 0.
\end{equation}
 
We are not able to solve 
(\ref{eq:tmatrix}) analytically for arbitrary 
configuration of pairs of kinks. 
However it is not so difficult to solve it numerically.
We can also easily obtain eigenfunctions after having energy eigenvalues
by using TM method.
It is straightforward to extend the above formalism for nonvanishing $m_0$.

Let us introduce the constant imaginary vector potential $g$.
Effect of the imaginary vector potential was discussed by Hatano and 
Nelson\cite{HN}. 
Let us denote the eigenfunction of energy $E$ for $g=0$ as $\Psi_{0}(x)$, 
and suppose the shape of $\Psi_{0}(x)$ as
\begin{equation}
 \Psi_{0}(x) \cong \exp \left(  -\frac{|x-x_{c}|}{\xi_{0}} \right),
\end{equation}    
where $\xi_{0}$ is the localization length
and $x_{c}$ is the center of this localized state.
When we turn on the constant imaginary vector potential $g$, 
the eigenfunction is obtained from
$\Psi_{0}(x)$ by the ``imaginary" gauge transformation,
\begin{equation}
 \Psi(x) \cong \exp \left( -\frac{|x-x_{c}|}{\xi_{0}} -\ g\ (x-x_{c})
 \right). 
\label{psix}  
\end{equation}
This means that the localization length of this eigenfunction is
\begin{eqnarray}
 \xi_{g} & = &  \frac{\xi_{0}}{1+g\xi_{0}} \ ( x>x_{c} ) \nonumber \\
 \xi_{g} & = &  \frac{\xi_{0}}{1-g\xi_{0}} \ ( x<x_{c} ).
\end{eqnarray}
At the point $g=1/\xi_{0}$, the localization length for $x<x_{c}$ diverges. 
 If the imaginary vector potential $g$ gets larger than this value, 
$\xi_{g}$ for $x<x_{c}$ becomes negative, 
and the eigenfunction cannot exist at the same energy since it
cannot satisfy periodic boundary condition (Fig.2).

So if localized eigenstate with energy $E$ disappears at $g=g_{c}$ as $g$ is
increased, then the localization length of the eigenstate $\Psi_{0}(x)$
is $1/g_{c}$.
Actually it is shown that for $g>g_c$, energy eigenvalue of the state
has an imaginary part and the state is extended as we shall see
shortly\cite{HN}.

The TMM can be easily extended for the case
of nonvanishing $g$.
We obtain energy eigenvalues and eigenfunctions numerically.
In actual calculation the system size is finite, and the energy
eigenvalues of the wavefunctions for $g=0$ and $g \neq 0$ are slightly
different as we mentioned above.
We must trace the changes of energy eigenvalues of states for
nonzero $g$ on the two dimensional $(E-g)$ plane 
and find the end point of the locus (Fig.3).

In the numerical calculation, we first find the energy eigenstate for
 $g=0$. We denote its energy eigenvalue as $E_{0}$. 
 Then we increase $g$ and search the eigenstate in the region 
 $E_{0}-\delta E < E <  E_{0}+\delta E$. (We set $\delta E$ 
 very small.)
  Let us suppose that the eigenstate
 vanishes in this energy region at the value $g=g_{1}$. We denote 
 the energy eigenvalue of the eigenstate for $g=g_{1}$ as $E_{g_{1}}$.
 We change the energy region for searching eigenstate 
 as $E_{g_{1}}-\delta E < E < E_{g_{1}}+\delta E$
 and let $g$ larger than $g_{1}$. We repeat this procedure. 
If we cannot find energy eigenvalue at the value $g$ larger than
 $g_{c}$, we judge that the localization length of the eigenstate
 for $g=0$ is $1/g_{c}$.
\begin{figure}
\begin{center}
\unitlength=1cm
\begin{picture}(15,6.5)
\centerline{
\epsfysize=7cm
\epsfbox{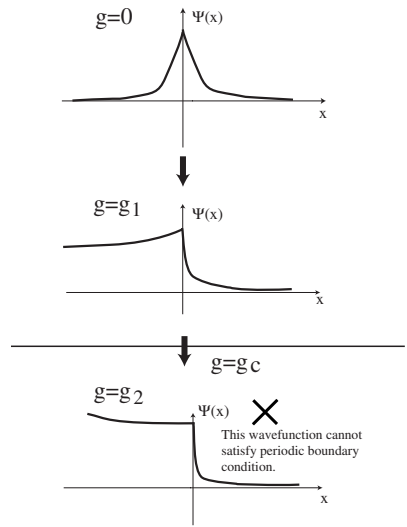}
}
\end{picture}
\end{center}
\caption{\small The imaginary vector potential method}
\label{fig:imvector}
\end{figure}
\begin{figure}
\begin{center}
\unitlength=1cm
\begin{picture}(15,6.5)
\centerline{
\epsfysize=6cm
\epsfbox{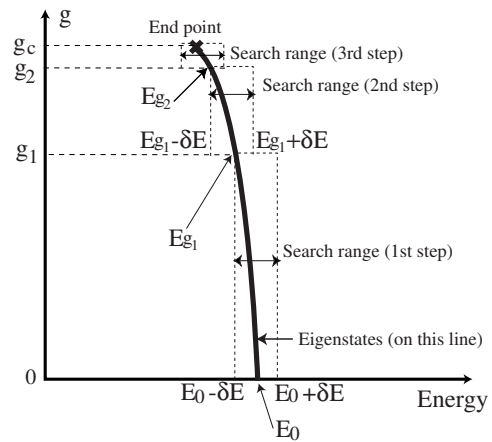}
}
\end{picture}
\end{center}
\caption{\small The procedure of imaginary vector potential method
 in the numerical calculation}
\label{fig:imvector2}
\end{figure}
First of all, we show that the delocalization transition actually
occurs at finite $g$.
In Fig.4, we show the wave functions of a low-lying state
in vanishing and nonvanishing imaginary vector potential.
For $g=0$, the state is obviously localized whereas at $g=0.03$
the state becomes extended and the energy eigenvalue has an imaginary
part.
As discussed in Ref.\cite{HN}, the density distribution of
a particle is given by $|\Psi(x,-g)\Psi(x,g)|$ where $\Psi(x,-g)$
is equal to the left eigenfunction.
\begin{figure}
\label{fig:wavef}
\begin{center}
\unitlength=1cm
\vspace{2mm}
\begin{picture}(15,3)
\centerline{
\epsfysize=3.2cm
\epsfbox{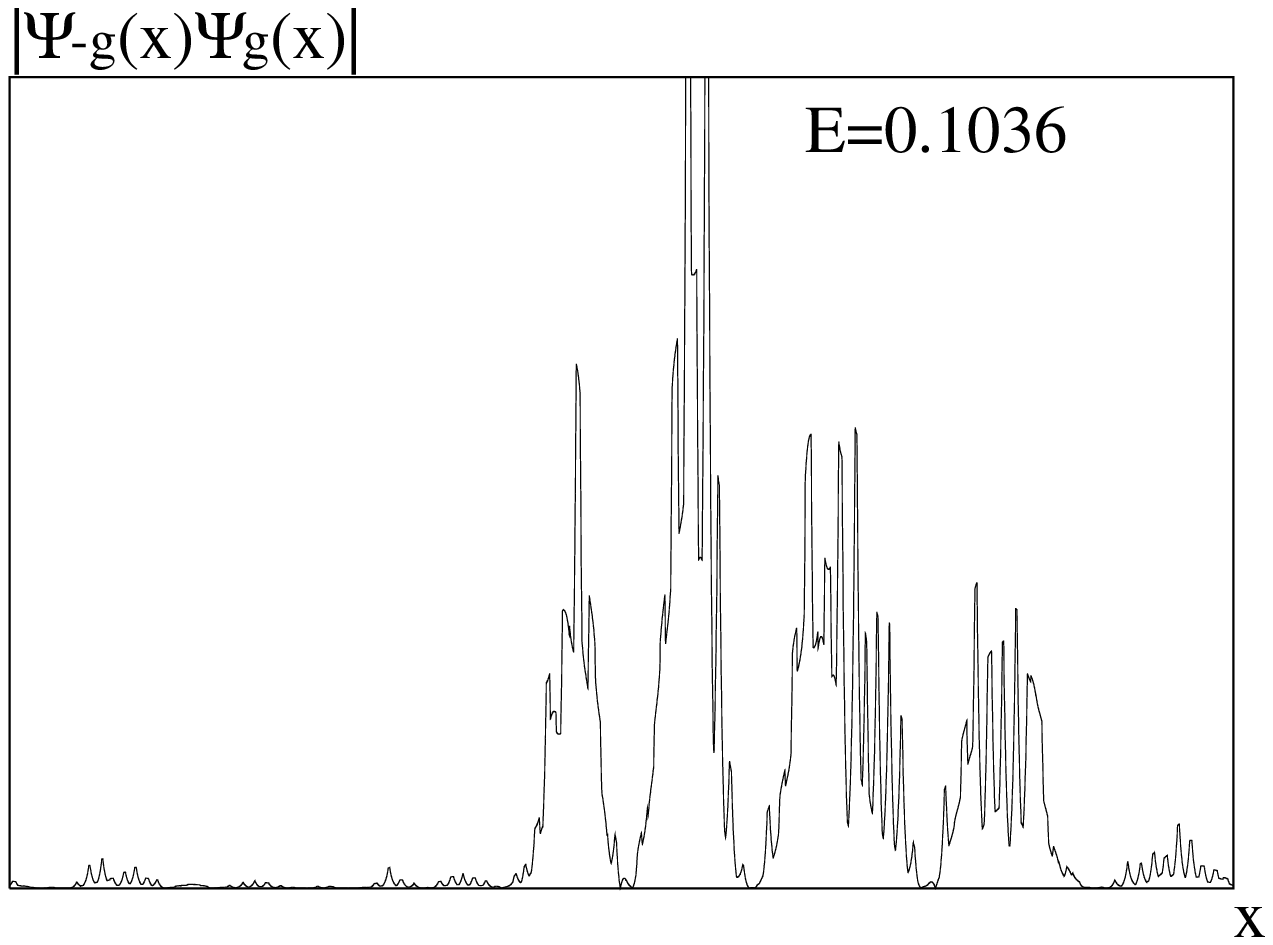}
}
\end{picture}
\begin{picture}(15,3)
\centerline{
\epsfysize=3.2cm
\epsfbox{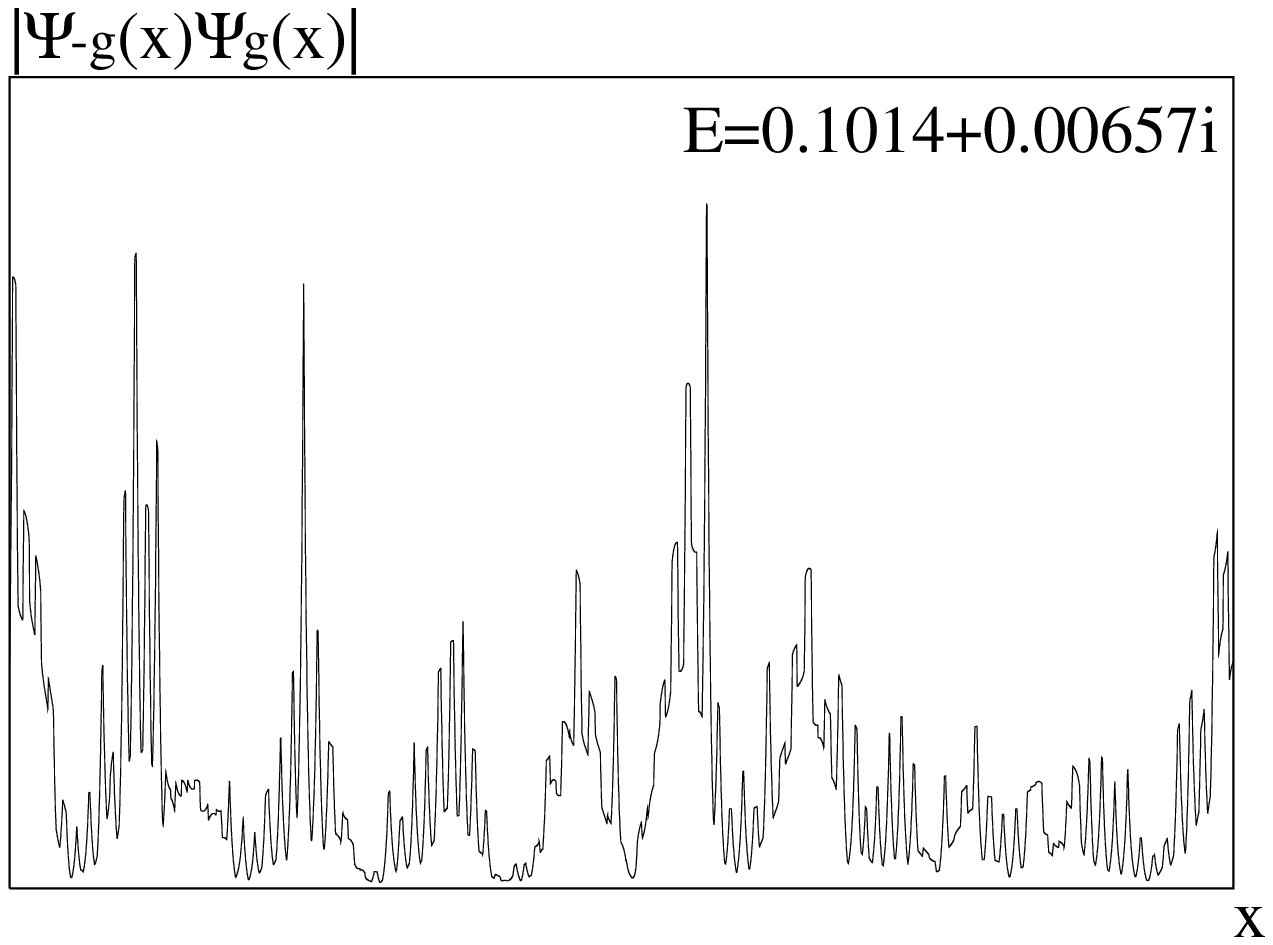}
}
\end{picture}
\vspace{0mm}
\caption{An example of the localized and the extended wave functions
in vanishing and nonvanishing imaginary vector potential $g=0$ and $g=0.03$.
The state of $g=0.03$ has a complex energy eigenvalue.}
\end{center}
\end{figure}

Let us turn to the localization length.
For the white-noise case $ [m(x)m(y)]_{\it{ens}} = A\ \delta(x-y)$, 
``typical" localization length or the inverse of the Lyapunov exponent
was obtained as\cite{Comtet,Theodorou,Eggarter}
\begin{equation}
 \xi_t(E) = |\ln E/2A|.
\label{eq:localize1} 
\end{equation}
Numerically the typical localization length $\xi_t(E)$ is obtained by
averaging over localization lengths of {\em all} eigenstates with energy $E$. 
On the other hand,
Balents and Fisher calculated the averaged Green function 
and obtained the mean
localization length from the spatial decay of the Green function\cite{BF}.
The result is
\begin{equation}
 \xi_m(E) = |\ln E/2A|^{2}.
\label{eq:localize2}
\end{equation}
Case of nonlocally correlated random mass was studied in 
Refs.\cite{IK,IK2} and $\xi_m(E)$ is obtained as a function of
$\tilde{\lambda}$ in Eq.(\ref{disordercor}).

By numerical calculation we obtain both the typical and mean
localization lengths.
As we mentioned above, the typical localization length is the average 
over all solutions of the Schr$\ddot{\mbox{o}}$dinger
equation ({\ref{eq:schroedinger1}),
whereas the mean localization length is determined by the states
which make dominant contributions to the Green function, i.e., which
have large localization length. 

The result of the numerical calculation of the typical
localization length of the white-noise case is given in 
Fig.5.  
We show the ratio of the numerical results to the analytical expression
in Eq.(\ref{eq:localize1}) in order to compare these two results. 
Therefore if the energy dependences of the localization
lengths obtained numerically and analytically are the same, 
this ratio should be constant. 
In Fig.5, 
the ratio seems constant over the whole range of $E$.   

 In Fig.6, we show the numerical results
 of the ``mean" localization length. 
 Here we use the solutions to the Schr$\ddot{\mbox{o}}$dinger equation
 which have long localization length. 
 More precisely, ``large"
 localization length $\xi$ means the one which satisfies   
 $\xi  > $ (the ``typical" localization length)+ 1.0 $\sigma$ 
 in each energy slice.
 From Fig.6, we conclude that the energy dependence 
 of the mean localization length obtained numerically is in agreement with 
 Eq.(\ref{eq:localize2}).  

\begin{figure}
\label{fig:fit1}
\begin{center}
\unitlength=1cm

\begin{picture}(15,3)
\centerline{
\epsfysize=4cm
\vspace{0mm}
\epsfbox{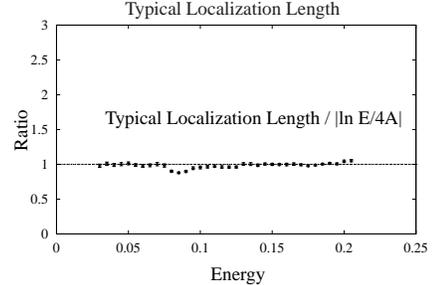}
}
\end{picture}
\vspace{0mm}
\caption{The comparison between the analytical and the numerical
results of the typical localization length for the (almost) white-noise case,
{\it i.e.}, small-$\lambda$ case:
In the numerical calculation, we set $L$(system size)$=50$, 
$A=1/2$, $\lambda=1/60$ in Eq.(\ref{disordercor}),
and the energy slice $\Delta E=0.01$. 
The ratio is normalized at $E=0.15$, and 
this result is averaged over about 15000 eigenstates.
The analytical and the numerical results are in good agreement.}

\end{center}
\end{figure}

\begin{figure}
\label{fig:fit2}
\begin{center}
\unitlength=1cm
 
\begin{picture}(15,3)
\centerline{
\epsfysize=4cm
\epsfbox{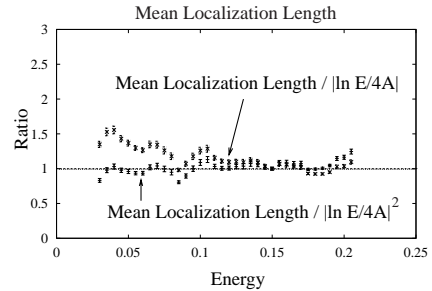}
}
\end{picture}
\vspace{0mm}
\caption{The comparison between the analytical and the numerical
results of the mean localization length:
Here we picked up the eigenstates which have large localization
length. (See the text.) 
We show the ratios of the numerical calculation both to 
Eqs.(\ref{eq:localize1}) and (\ref{eq:localize2}).
We set $L=50$, $A=1/2$ and the energy slice $\Delta E=0.02$.
 The ratios are normalized at $E=0.15$.}
\end{center}
\end{figure}

We shall turn to the case of the nonlocally-correlated disorder. 
In the white-noise limit ($\tilde{\lambda}=0$ case in Eq.(\ref{disordercor})), 
the localization lengths 
diverge only at $E=0$, that is, extended states exist only at $E=0$. 
If we let $\tilde{\lambda} > 0 $, the random mass becomes
nonlocally-correlated, and the critical energy or the mobility edge
at which the delocalization transition 
occurs may change.

We investigate the ``typical" and ``mean" localization lengths 
in the case of nonvanishing $\tilde{\lambda}$'s. 
The behaviour of the ``typical" and ``mean"
localization lengths obtained as in the white-noise case are given 
in Figs.7 and 8.
It seems that there is {\em no} $\tilde{\lambda}$-dependence
in the typical localization length.
On the other hand, Fig.8 shows 
that the mean localization length has 
a small but finite dependence on $\tilde{\lambda}$.
 
From the above calculations, we conclude that the effect of the 
short-range correlations in disorders is 
not so large. 
Especially the result indicates that the delocalization transition 
occurs at $E<0.03$. (If the mobility edge exists at $E_c>0$, 
the ratio in Fig.7 or 8 must diverge at $E_c$.) 
The delocalization transition probably occurs at $E=0$. 

\begin{figure}
\label{fig:fit3}
\begin{center}
\unitlength=1cm

\begin{picture}(15,3)
\centerline{
\epsfysize=4cm
\epsfbox{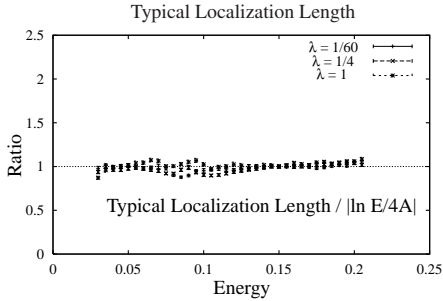}
}
\end{picture}
\vspace{0mm}
\caption{The behaviour of the localization length in the case of nonvanishing 
 $\lambda$'s: 
 Here we show the ratios of the typical localization lengths to $|\ln E/4A|$
 (Eq.(\ref{eq:localize1})).
 We set $L=50$, $A=1/2$ and the energy slice $\Delta E=0.01$. 
 The ratio is normalized at $E=0.15$.}
\end{center}
\end{figure}

\begin{figure}
\label{fig:fit4}
\begin{center}
\unitlength=1cm

\begin{picture}(15,3)
\centerline{
\epsfysize=4cm
\epsfbox{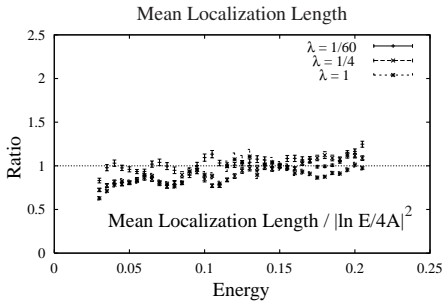}
}
\end{picture}
\vspace{0mm}
\caption{The behaviour of the localization length in the case of nonvanishing 
 $\lambda$'s: 
 We show the ratios of the mean localization lengths to $|\ln E/4A|^2$
 (Eq.(\ref{eq:localize2})).
 We set $L=50$, $A=1/2$ and the energy slice $\Delta E=0.02$.
 The ratio is normalized at $E=0.15$.}
\end{center}
\end{figure}

In the previous paper\cite{IK2} we calculated the localization length 
for the random mass with the short-range correlation (\ref{disordercor}).
We obtained the ``mean'' localization length to the 1st order of 
$\tilde{\lambda}$ by means of the Green function method.
 The result is
\begin{equation}
  \xi(E) = \frac{1}{A} \Big( \frac{\ln |\frac{E}{2A}|^{2}}{\pi^{2}}
+A \tilde{\lambda} \frac{4|\ln \frac{E}{2A}|}{\pi^{2}}\Big)
+O(\tilde{\lambda}^2).
\label{eq:localize3}
\end{equation}
In Fig.9 we show 
the ratios of the numerical result to the analytical
calculation up to the 0th and the 1st order of $\tilde{\lambda}$.
This shows that the analytical result with the 1st order correction 
of $\tilde{\lambda}$ is in better agreement with the numerical result, 
but the correction is small.

\begin{figure}
\label{fig:fit5}
\begin{center}
\unitlength=1cm

\begin{picture}(15,4)
\centerline{
\epsfysize=4cm
\epsfbox{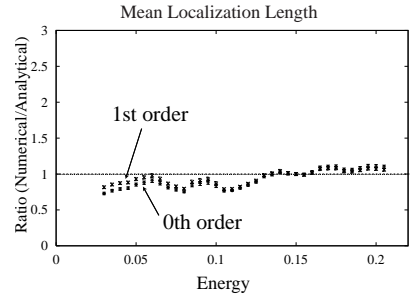}
}
\end{picture}
\vspace{0mm}
\caption{The comparison between the numerical result and the analytical one
 in the case of relatively large $\lambda$: 
 We show the ratios of the mean localization length calculated numerically 
 to the ones obtained analytically (Eq.(\ref{eq:localize3})).
 Here we used the analytical results
 of the 0th and the 1st order of $\lambda$. 
 We set $\lambda=1/4$, $A=1/2$ and $L=50$.
 The energy slice is $\Delta E=0.02$, and the ratio is normalized at
 $E=0.15$.}
\end{center}
\end{figure}

From the investigations given as far we can conclude that
the numerical methods used in this paper are reliable for
calculating the localization lengths.
It is very interesting to study the case of disorders with a long-range
correlation.  
We expect that a nontrivial mobility edge $E_c>0$ exists 
for a certain long-range correlated random mass.
Actually the one-dimensional Anderson model with
long-range correlated disorder was studied\cite{Lyra}, 
and it is shown that there exists a nontrivial mobility edge.

We can also calculate exponents of the multi-fractal 
scaling\cite{Huckestein} by the TMM\cite{TTIK}.
These problems are under study and results will be reported
in a future publication\cite{TI2}.

%%%%%%%%%%%%%%%%%%%%%%%%%%%%%%%%%%%%%%%%%%%%%%%%%%%%%%%%%%%%%%%%%%%%%%%%

\section{Gaussian-distributed distances between kinks: Correlation
length vs. localization lengths}
In subsequent papers\cite{TI,TI2} we shall study the localization length
with long-range correlated disorder with the TMM and IVPM.
Before that study, we shall consider the case of Gaussian distribution
for distances between kinks in this section.
That is, the probability distribution
 function $P_{{\rm G}}$ for distance $l$ is given by
 the Gaussian distribution,
\begin{equation}
P_{{\rm G}}(l)=\frac1{\sqrt{2\pi}\sigma}\exp\{-(l-\mu)^2/2\sigma^2\},
\end{equation}
where $\mu$ is the mean distance and $\sigma$ is the standard deviation.
If $\sigma$
 is small, eigenstates at low energy (eigenstates which have
 a small number of nodes) spread over whole system, and a
 mobility edge may exist at a nonzero energy.
 Physically, the
mean distance corresponds to the kink density ( it can be viewed as 
impurity density in the related models such as 
the random hopping tight binding model, spin chain, etc.)
and standard deviation controls the degree of randomness.
This is in sharp contrast to $P(l)$ in Eq.(\ref{Pl}) in which
only one parameter $\tilde{\lambda}$ exists.

 We first show the relation between localization lengths
 of eigenstates and the deviation $\sigma$. 
 The result is shown in Fig.10.
There are several peaks of localization length
 in these figures with small $\sigma$.
 We know that the energy spectrum is
 discrete and all states become extended in the case of periodic
 $m(x)$, i.e., the limit $\sigma \rightarrow 0$,
  as dictated by the Bloch theorem.
 For small but finite $\sigma$,
 we expect that energy spectrum is not discrete and
 eigenvalues spread within narrow energy band.
 We also expect that localization lengths
 of states are very large, but finite for nonvanishing $\sigma$.

\begin{figure}
\begin{center}
\unitlength=1cm

\begin{picture}(10,6.5)
\put(-2,4){
\centerline{
\epsfysize=2.8cm
\epsfbox{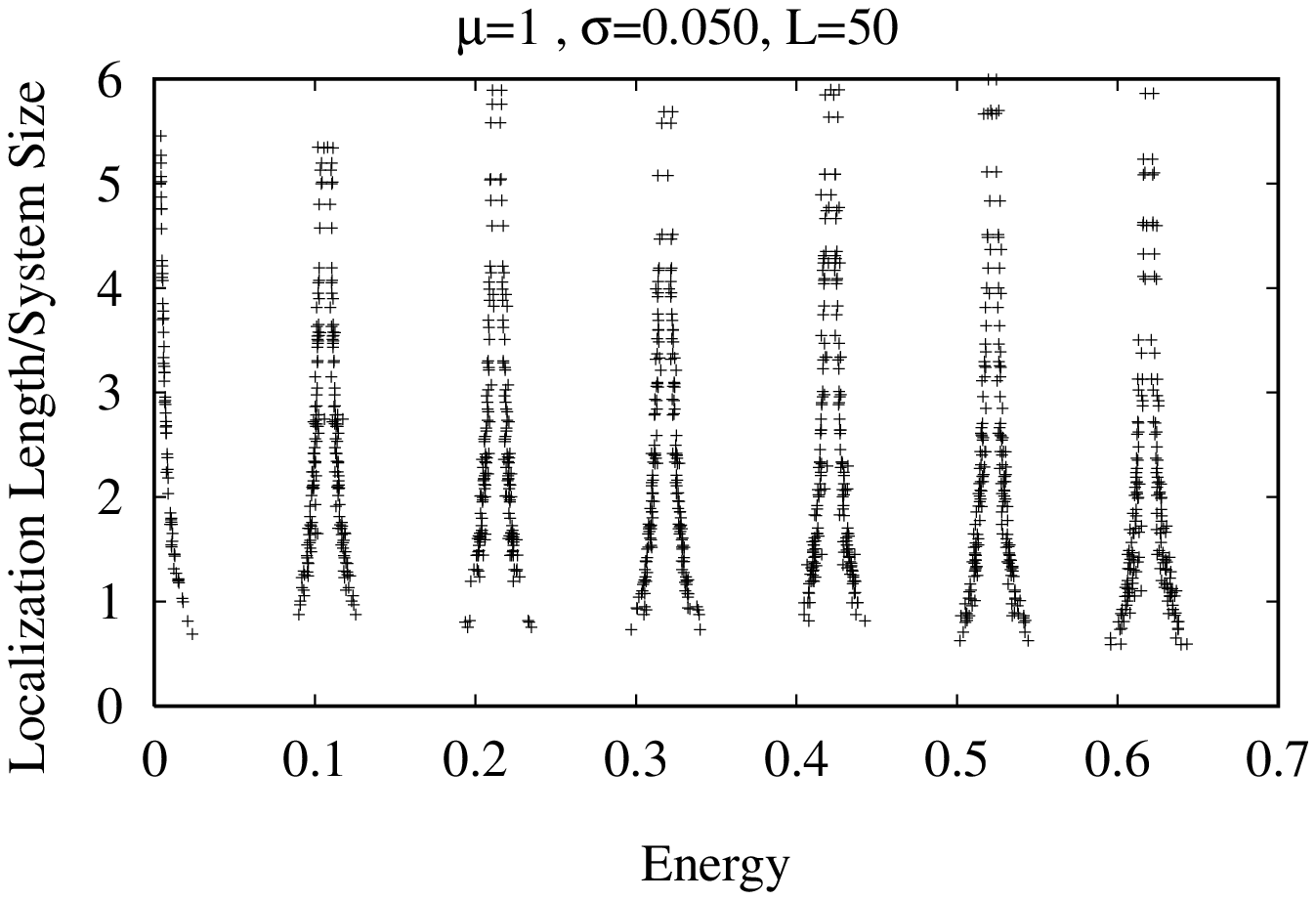}
}}
\put(2,4){
\centerline{
\epsfysize=2.8cm
\epsfbox{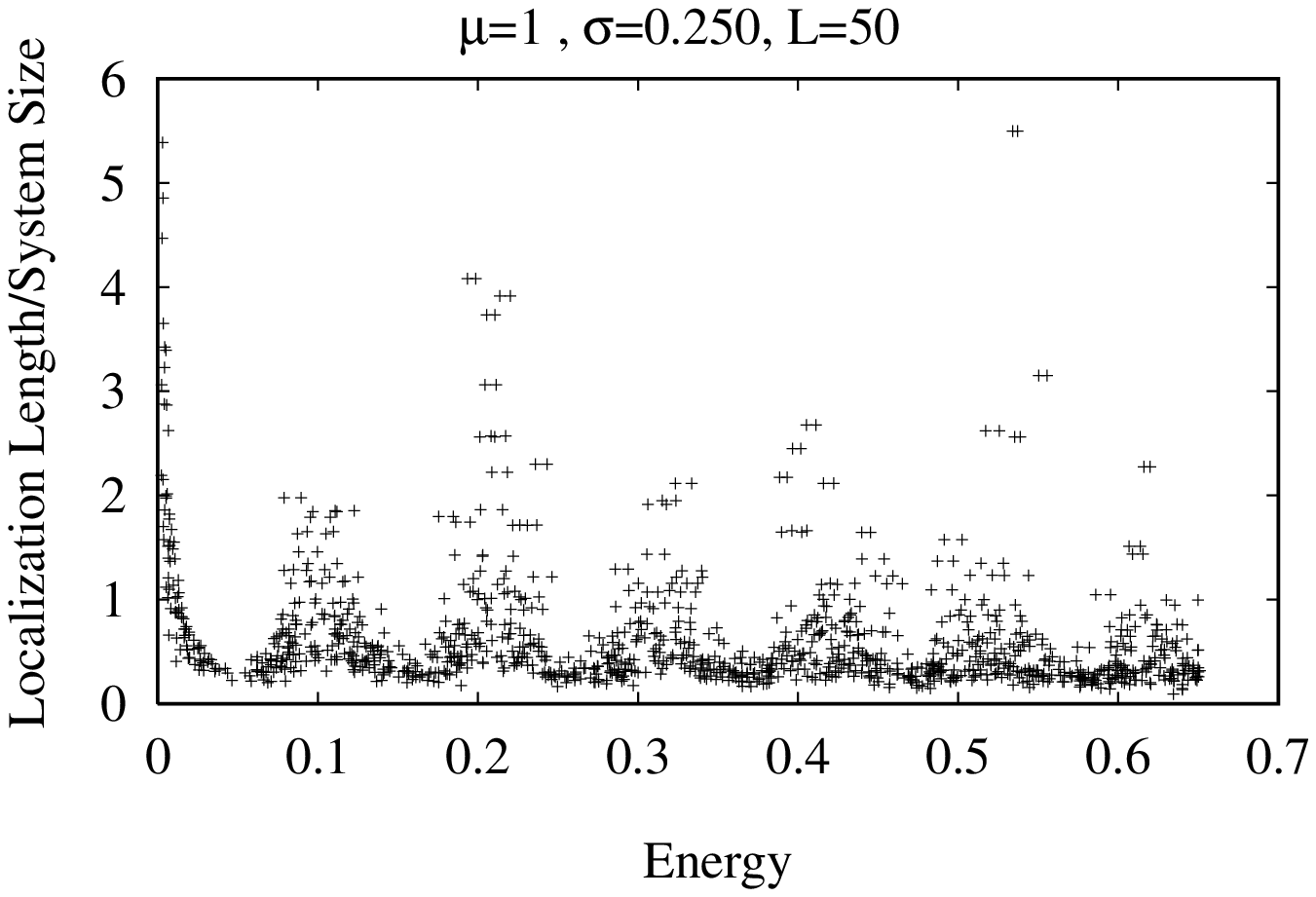}
}}
\put(-2,1){
\centerline{
\epsfysize=2.8cm
\epsfbox{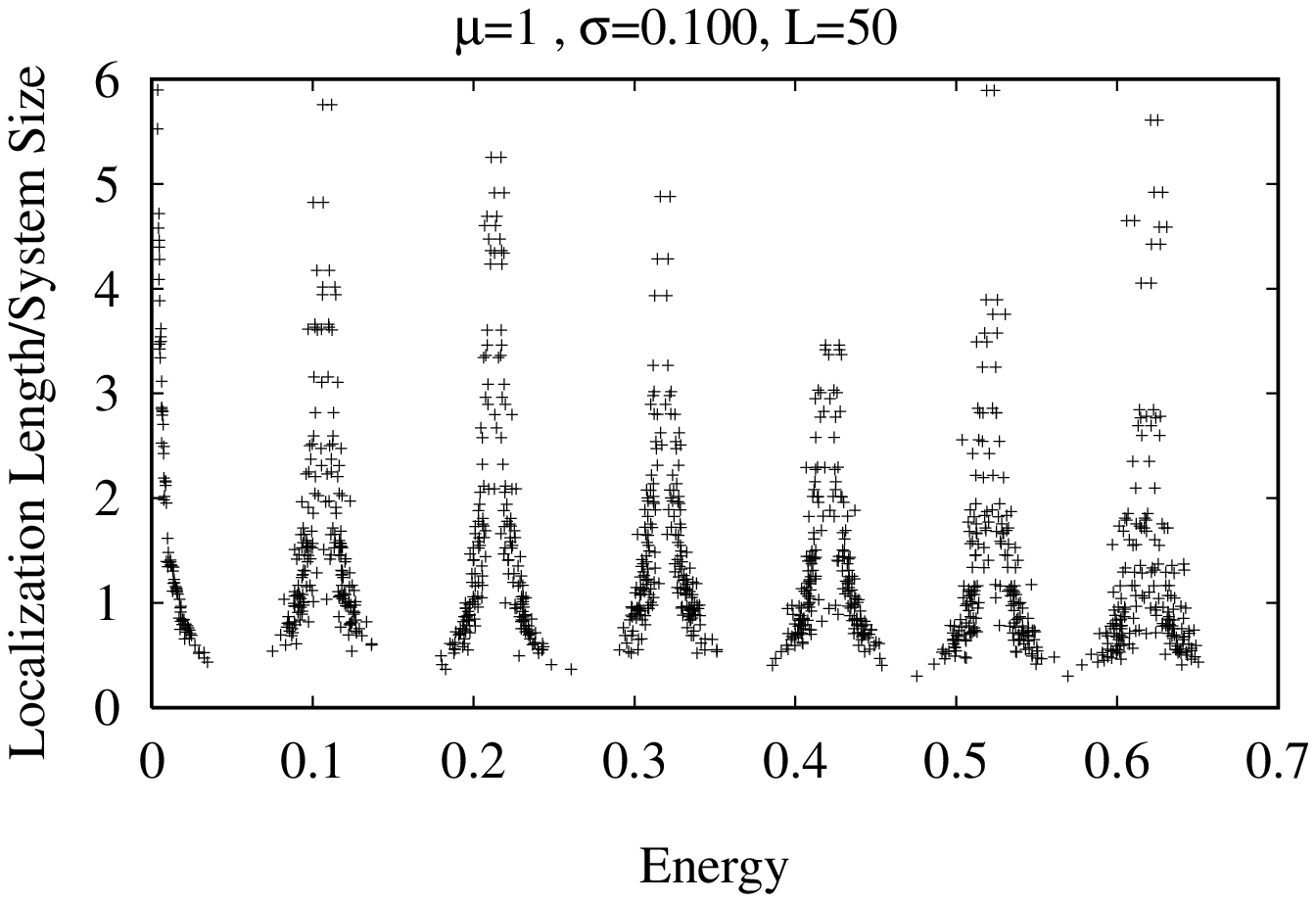}
}}
\put(2,1){
\centerline{
\epsfysize=2.8cm
\epsfbox{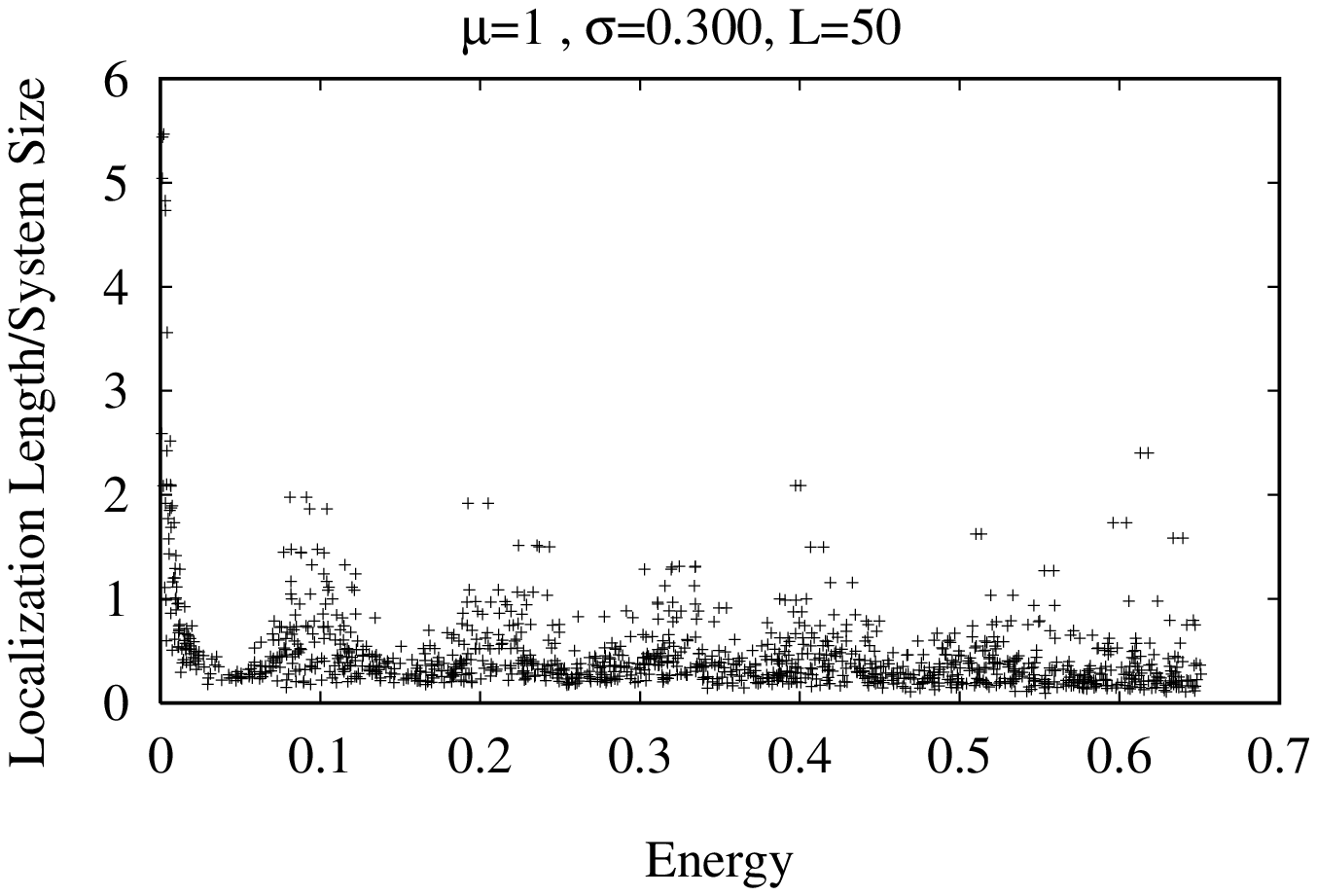}
}}
\put(-2,-2){
\centerline{
\epsfysize=2.8cm
\epsfbox{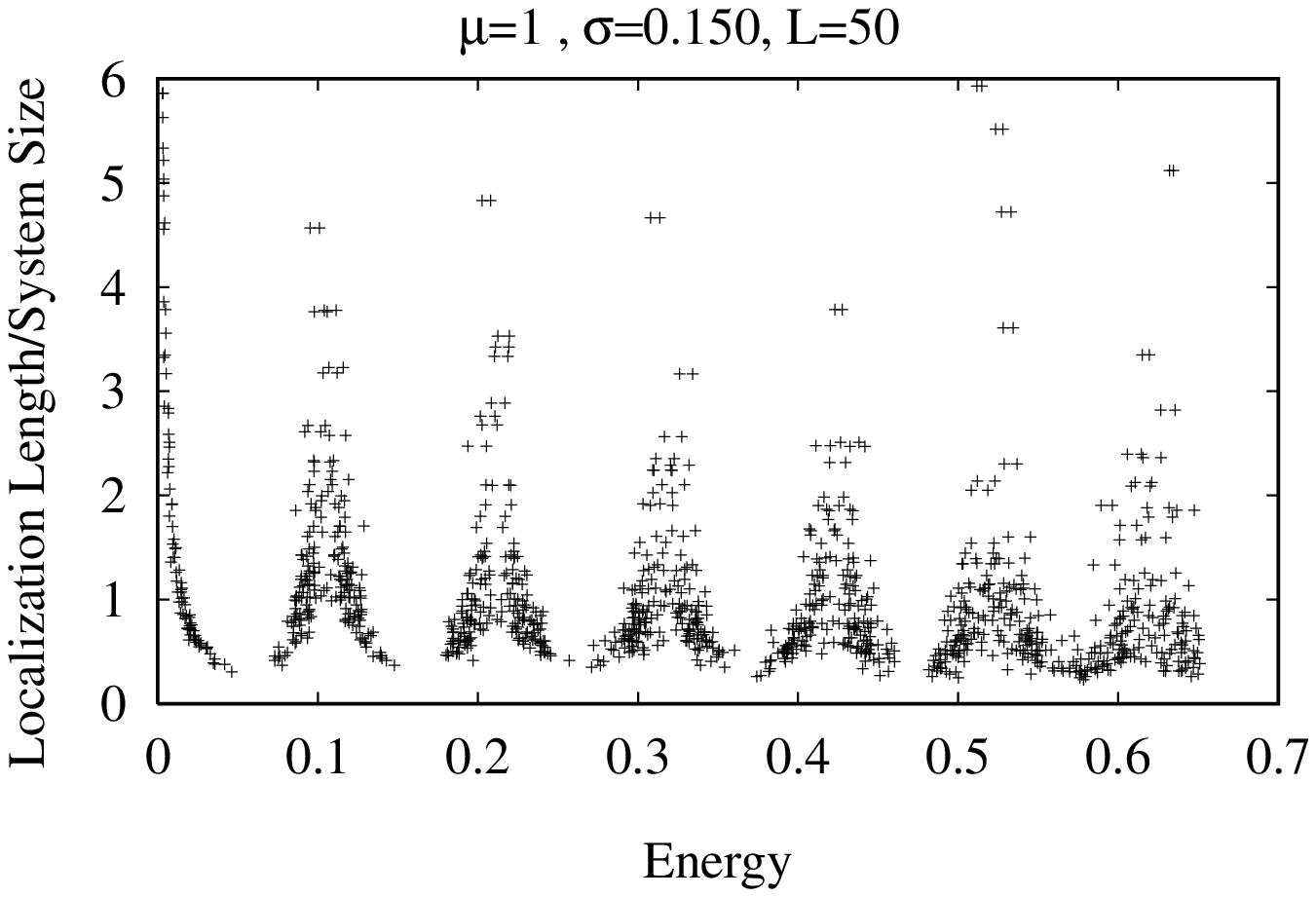}
}}
\put(2,-2){
\centerline{
\epsfysize=2.8cm
\epsfbox{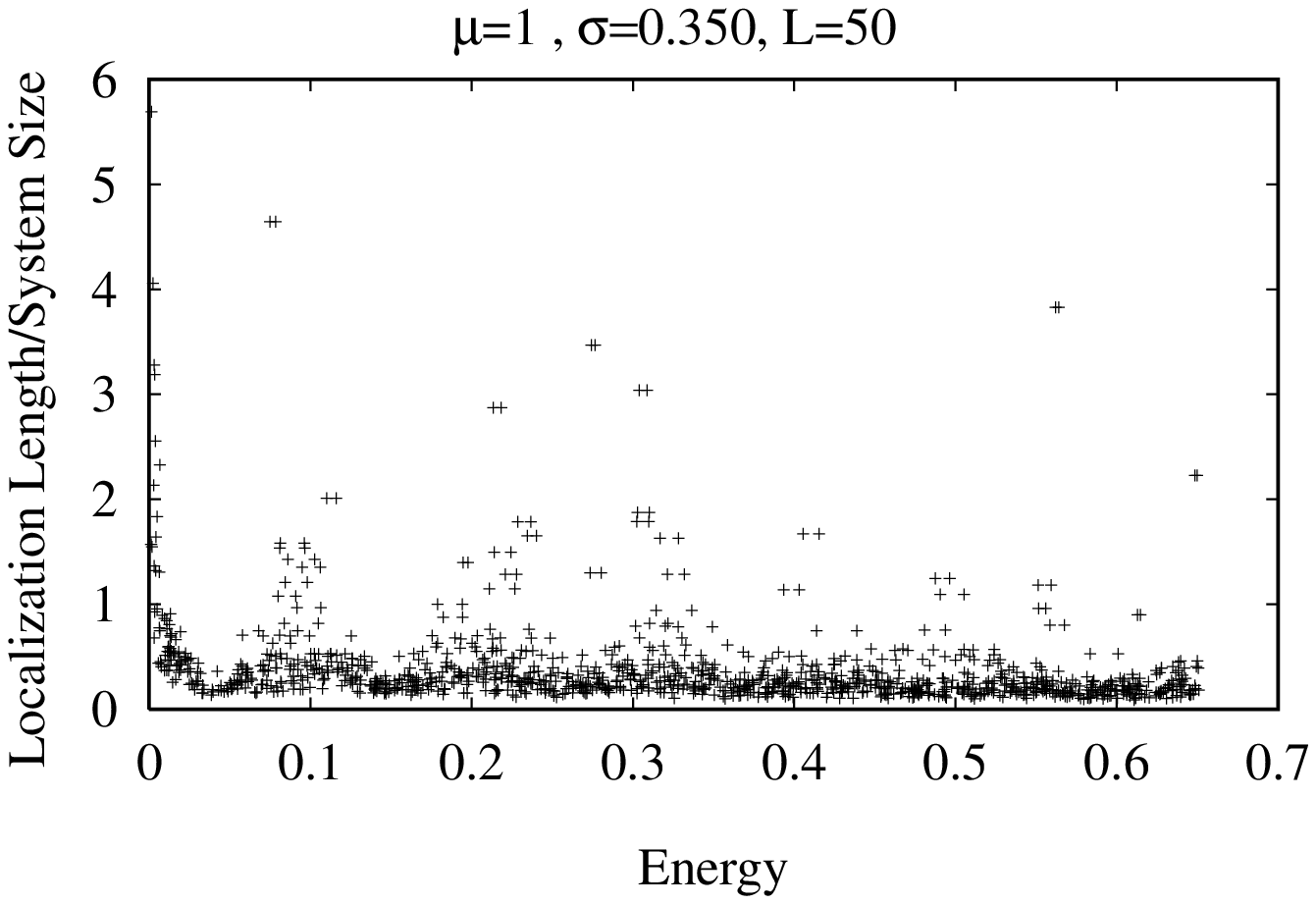}
}}
\put(-2,-5){
\centerline{
\epsfysize=2.8cm
\epsfbox{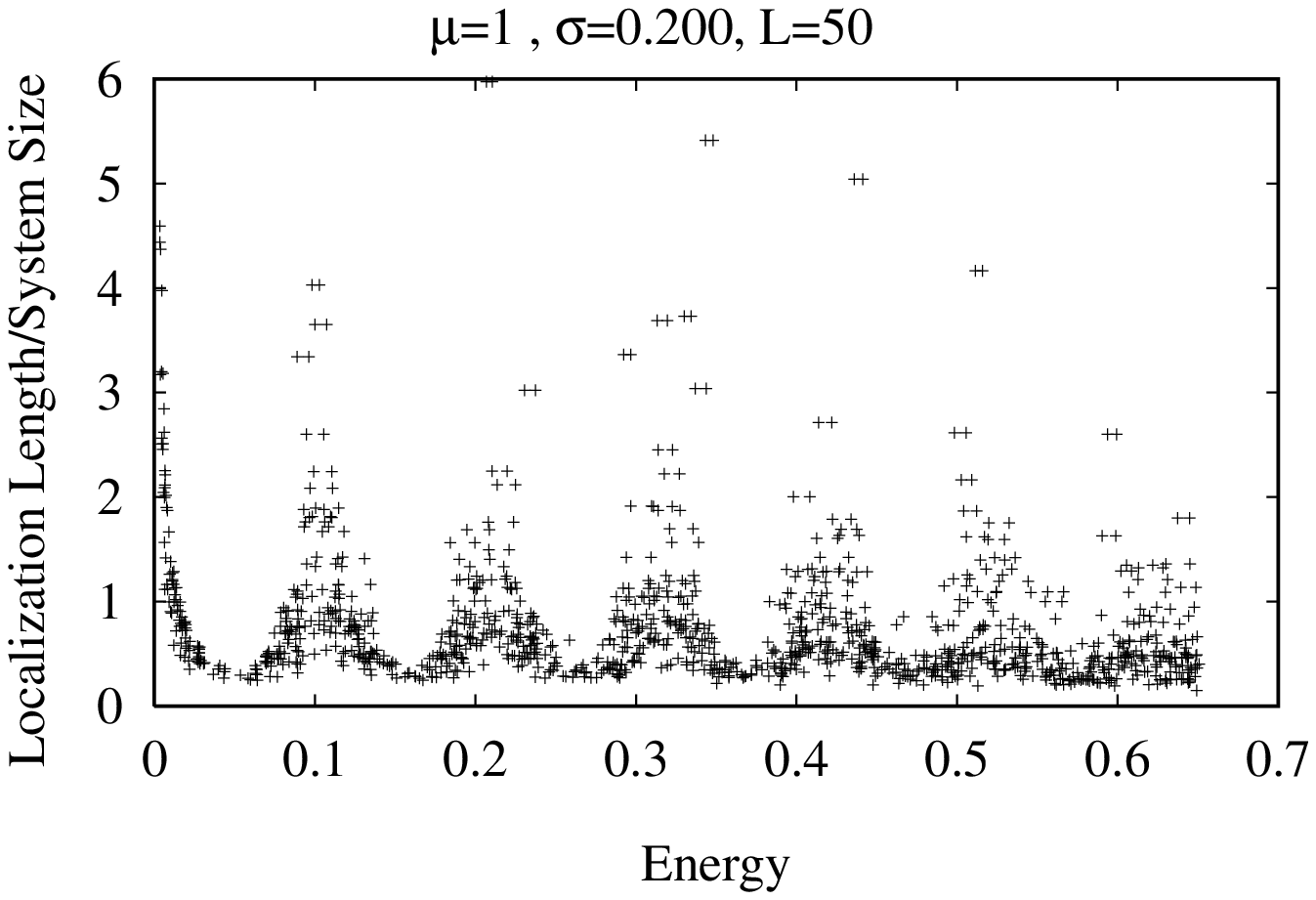}
}}
\put(2,-5){
\centerline{
\epsfysize=2.8cm
\epsfbox{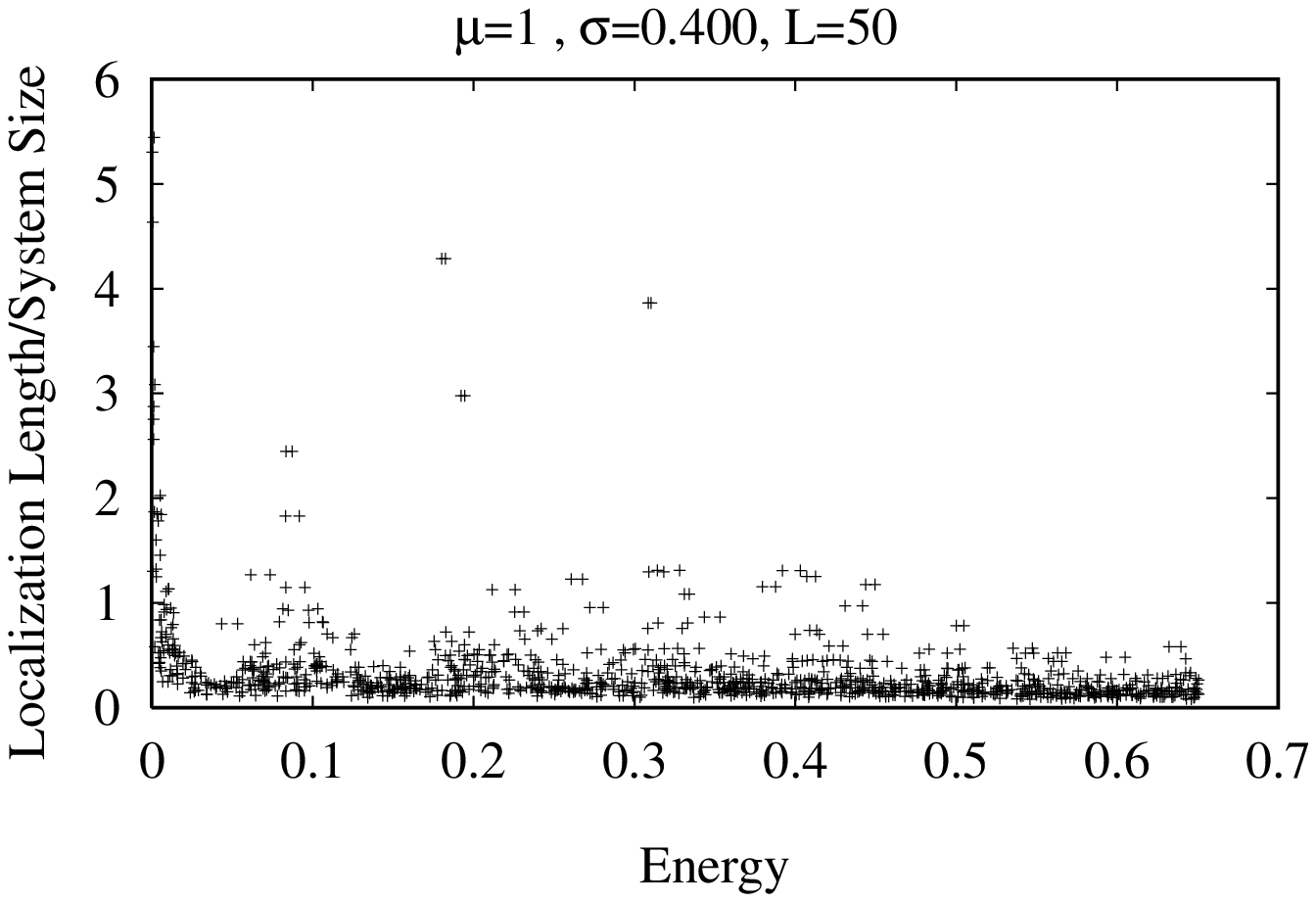}
}}
\end{picture}
\end{center}
\vspace{5cm}
\caption{Localization length and standard deviation of Gaussian distribution:
Here we set $L$(system size)$=50$.
In the case of small $\sigma$, the points 
concentrate around the eigenvalues of periodic system
 and they form several peaks.
 As we let $\sigma$ larger, the peaks disappear
 and these figures become similar to 
 the one in case of white-noise disorder.
(Extended state exists only at $E=0$ and other states localize.)}
\label{fig:explong1}
\end{figure}  

 In a system of {\em finite size}, we regard state as extended one
if the localization length is larger than the system
 size. We can see that localization lengths
 of several states become larger than the
 system size if we take $\sigma$ small. This means that
 some eigenstates become extended in finite size system. 

 If we let $\sigma$ much larger, these peaks of localization length
 get smaller and
 disappear except for the one at $E=0$. 
 We recall that the state at $E=0$ is ``isolated'' extended
 state in the system with white-noise correlated random mass.
 In the case of large $\sigma$, the random mass $m(x)$ becomes
 short-ranged and therefore all states except those close to $E=0$
 tend to localize. 
   
Next we show the two-point correlation function of random mass
 $[m(x)m(0)]_{\rm ens}$
 (Fig.11).
 It oscillates due to Gaussian distributed randomness. The
 period of this oscillation reflects 
 mean value of Gaussian distribution. 
 The value of envelope of $[m(x)m(0)]_{\rm ens}$ approaches to zero for
 larger $x$, and
 the envelope shows exponential decay, as we can see in
  Fig.11.
 This behaviour is the same as in the case of exponentially distributed kink
 distances. In the present case, however, we can obtain the system 
 with long-range 
 correlated disorder in which there are a {\it large} number of kinks.
 (In the case of exponential distribution $P(l)$ in Eq.(\ref{Pl}), 
 there are a small number of kinks
 in the system with large correlation length, and the distribution with 
 $P(l)$ is not sufficient
 for studies on practical systems like random spin chains.)
 The correlation length of
 random mass is very large and exceeds
 system size for small $\sigma$. 

\begin{figure}
\begin{center}
\unitlength=1cm

\begin{picture}(15,3.5)
\centerline{
\epsfysize=4cm
\epsfbox{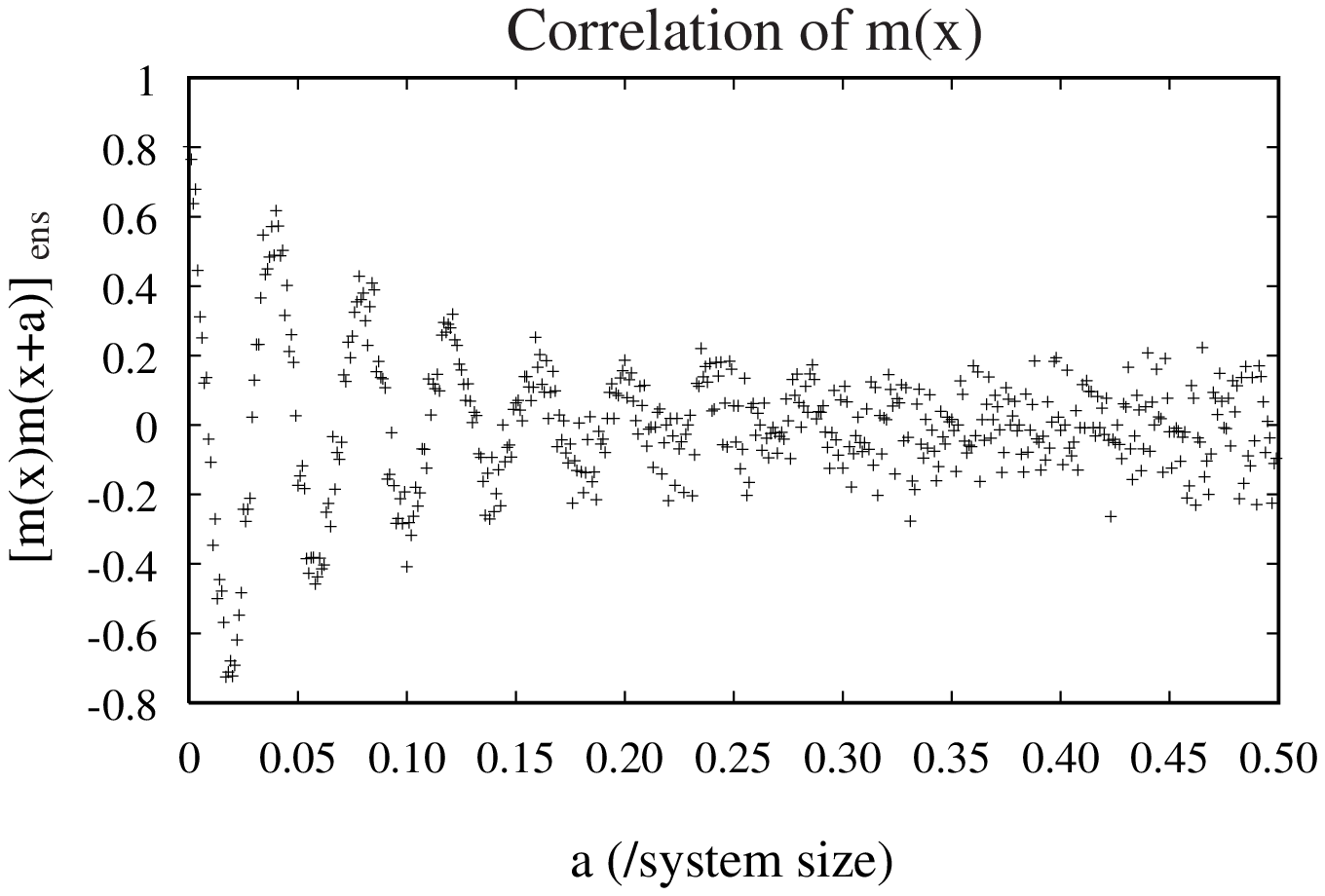}
}
\end{picture}
\begin{picture}(15,4)
\centerline{
\epsfysize=4cm
\epsfbox{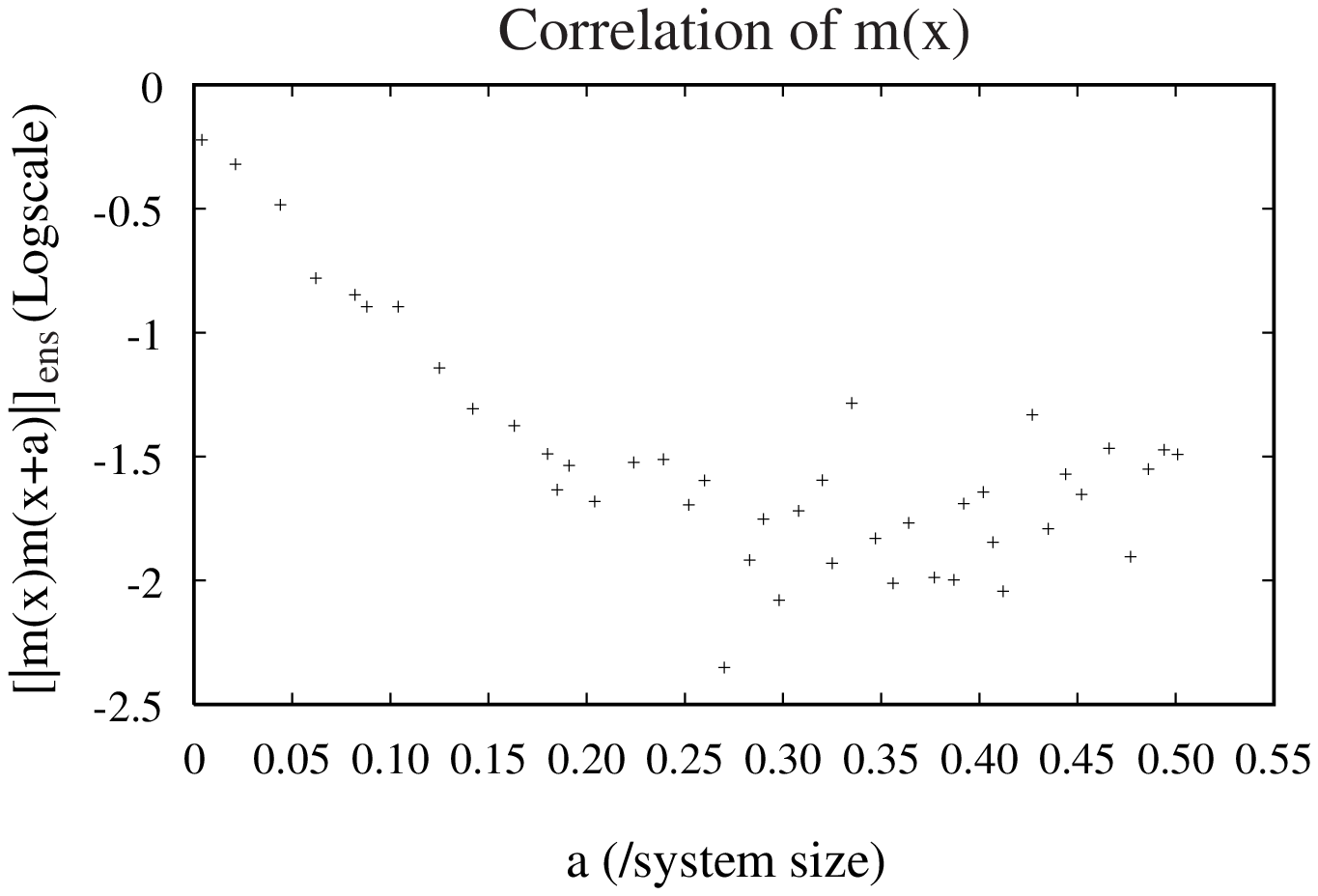}
}
\end{picture}
\end{center}
\vspace{5mm}
\caption{Correlation of random mass with Gaussian
distributed randomness: Here we set $\mu=1,\; \sigma=0.20$ and $L$(system 
size)$=50$. We show the behaviour of envelope in the logscale plot. 
 From this figure, we know that $m(x)$ is exponentially correlated in this case.}
\vspace{1cm}
\label{fig:expcor}
\end{figure}

 The relation between $\sigma$ and the correlation length $\lambda$
 (decay rate) of random mass is shown in Figs.12 and 13.
 ($\lambda$ is
 defined from the envelope of correlation of $m(x)$ as
 $[m(x)m(x+a)]_{\rm ens} \sim \exp (-a/\lambda) $.)
 From Fig.13, we conclude that these quantities are related as
 \begin{equation}
 \label{eq:lambdarel}
 \lambda = A \sigma^{-\nu}.
 \end{equation}
 In the case shown in Fig.13, $A$ and $\nu$
 are estimated as  
 \begin{eqnarray}
 && A=4.91 \times 10^{-3} ,\; \nu = 1.881 \; (L=50,\; \mu=1), \nonumber \\
 && A=2.67 \times 10^{-3} ,\; \nu = 1.837 \; (L=100,\; \mu=1), 
 \end{eqnarray}
 where $L$ is the system size.

\begin{figure}
\begin{center}
\unitlength=1cm

\begin{picture}(15,3.5)
\centerline{
\epsfysize=4cm
\epsfbox{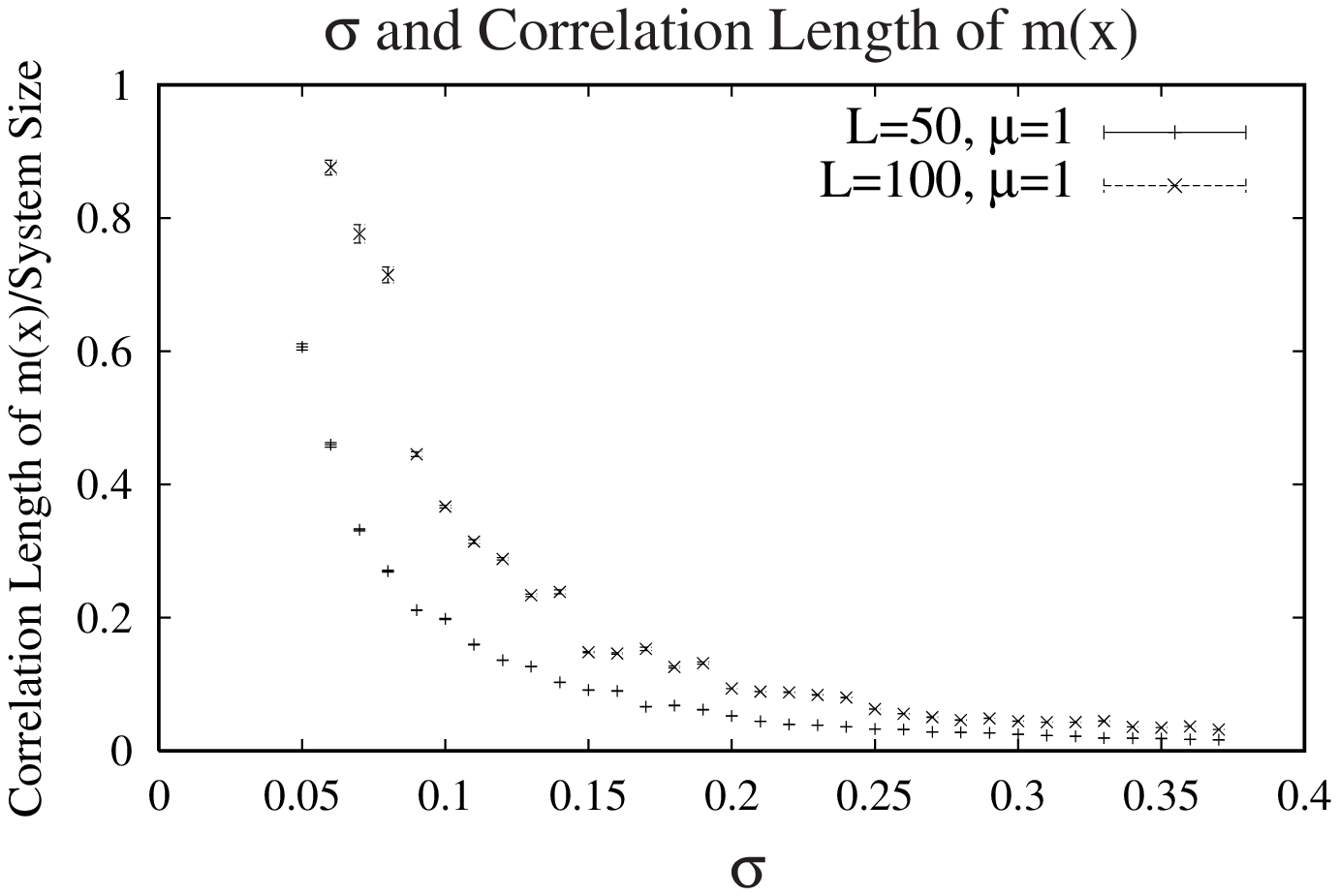}
}
\end{picture}
\vspace{0mm}
\caption{Relation between deviation $\sigma$ of Gaussian distribution and
 correlation length $\lambda$ of random mass(1): $\lambda$ is
 defined as $[m(x)m(x+a)]_{\rm ens} \sim \exp (-a/\lambda) $. In the case of small
 $\sigma$, correlation length of $m(x)$ is much larger than the system size.}
\label{fig:sigvsxi1}
\vspace{1cm}
\begin{picture}(15,4)
\centerline{
\epsfysize=4cm
\epsfbox{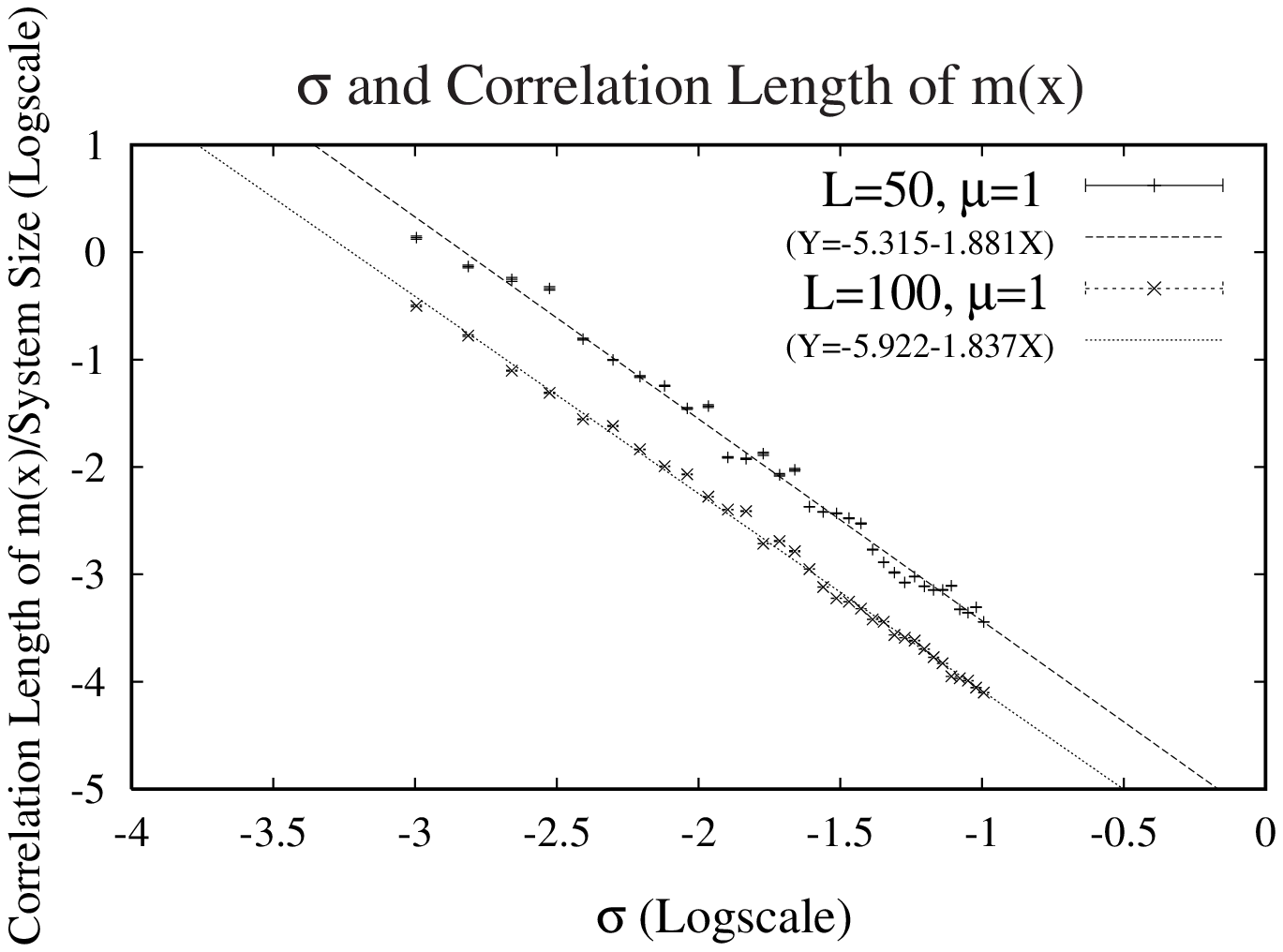}
}
\end{picture}
\end{center}
\vspace{0mm}
\caption{Relation between deviation $\sigma$ of Gaussian distribution and
 correlation length $\lambda$ of random mass(2) (Logscale {\em vs.}
 logscale plot):
 Points for fixed system size are almost on one line.}
\label{fig:sigvsxi2}
\end{figure}

 We observe the typical localization length of the states 
 at a fixed energy and also its dependence on system size.
 We have already seen that energy eigenvalues of the states concentrate
 around several energy values.
 In order to calculate typical localization length,
 we pick up the states around second peak from the lowest energy
 and average localization lengths of these states.
 (See Fig.10. In the actual calculation, we picked up
 the states around $E \simeq 0.1$ in the system with $L=50,\; \mu=1$, and  
 the ones around half(quarter) of this energy value in the
 double(four times) size system.) 

 The dependences 
 of typical localization length on $\sigma$ and system size are shown in
 Fig.14.
 We expect that the $\sigma$ dependence of localization length $\xi$
 can be fit by following function,
 \begin{equation}
 \label{eq:sigmarel}
 \xi = A \; | \sigma - \sigma_{c} |^{-\eta},
 \end{equation}
 where $\sigma_{c}$ is critical value and $\eta$ is
 critical exponent. Strictly speaking, this relation holds true only in
 infinite size system. We expect that we can assume this parameterization
 if the system size is large enough.  
 
 The result of fitting is shown in the Table 1.
 We conclude that the choice of this fitting function is valid
 from these $\chi^{2}$
 values and also from the fact that the critical exponents for different 
 system sizes are almost the same.

\begin{figure}
\begin{center}
\unitlength=1cm

\begin{picture}(15,4)
\centerline{
\epsfysize=4cm
\epsfbox{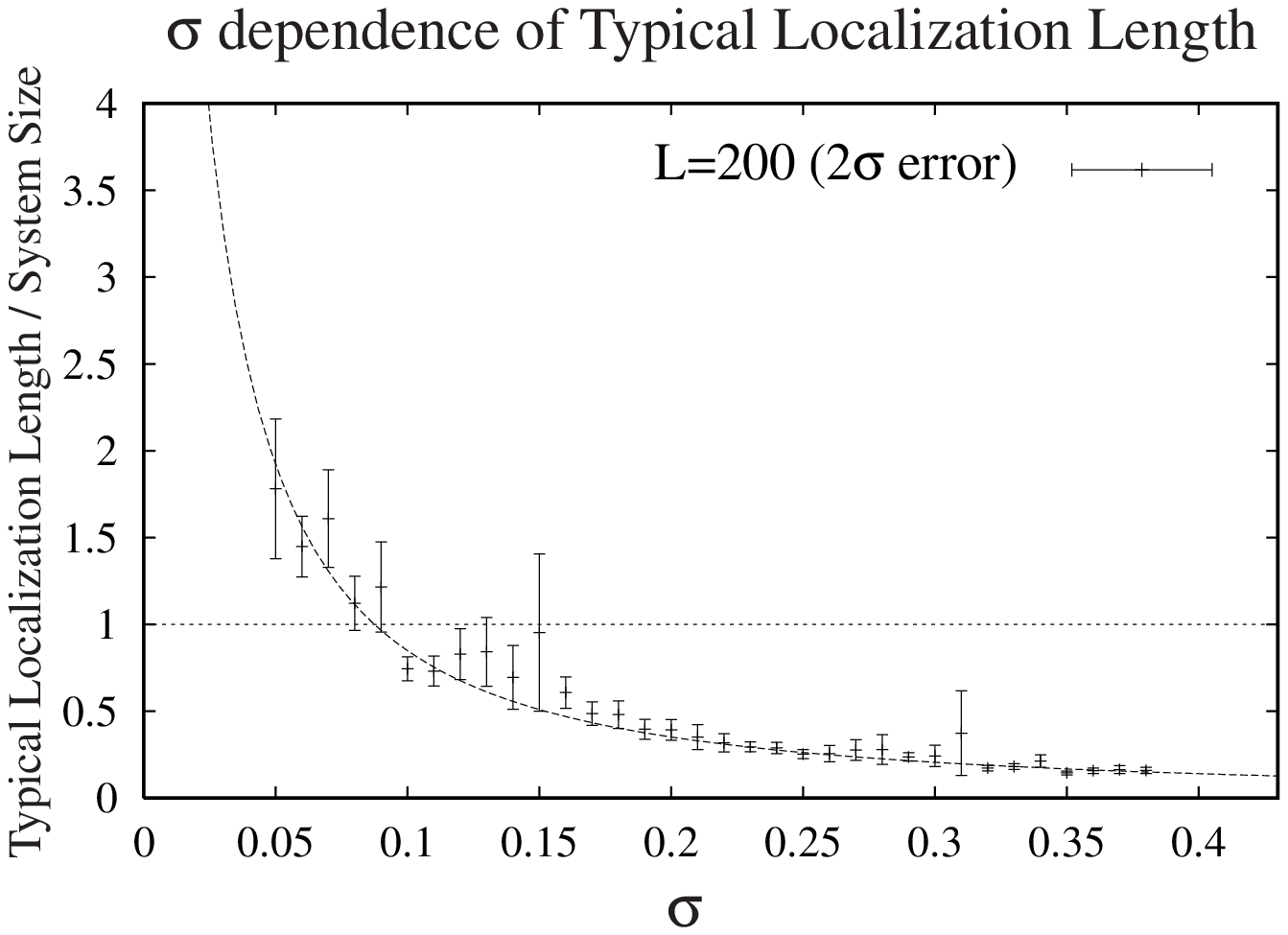}
}
\end{picture}
\vspace{0.5cm}
\begin{picture}(15,4)
\centerline{
\epsfysize=4cm
\epsfbox{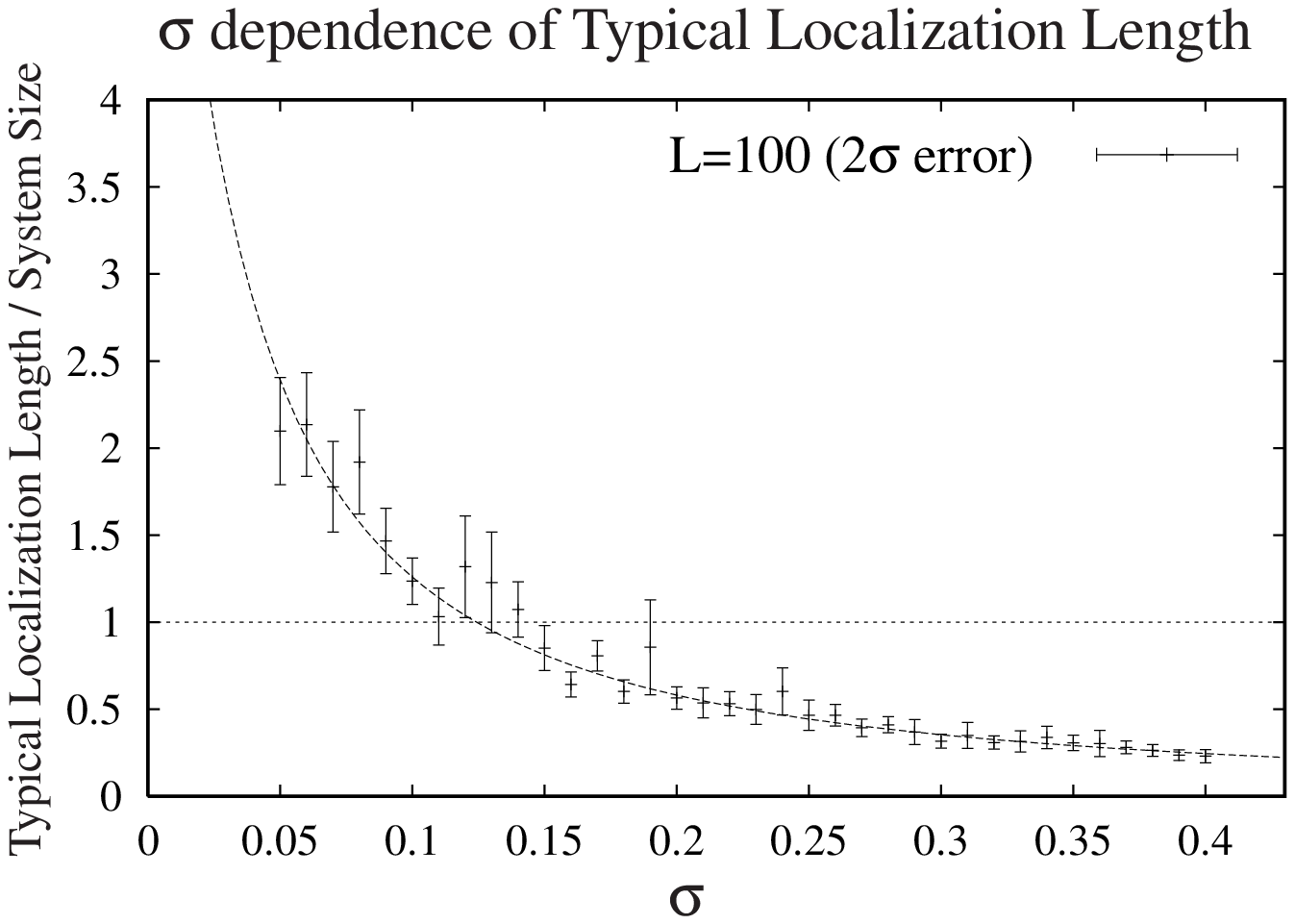}
}
\end{picture}
\vspace{0.5cm}
\begin{picture}(15,4)
\centerline{
\epsfysize=4cm
\epsfbox{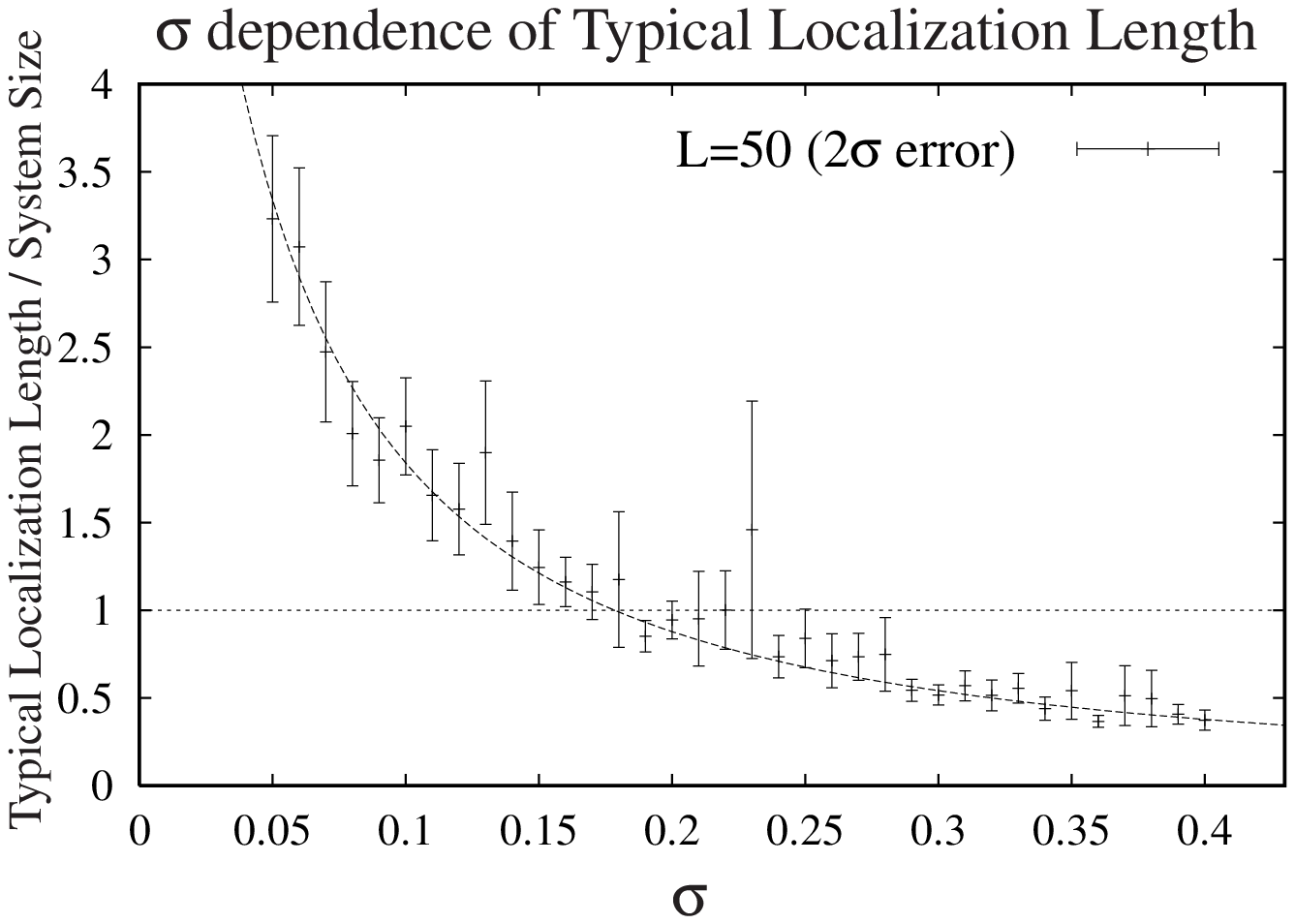}
}
\end{picture}
\end{center}
\vspace{0mm}
\caption{Typical localization length and deviation $\sigma$:
We see the $\sigma$ dependences of typical localization lengths for
 second peak in various size systems. Here we set $\mu$(mean distance)$=1$.
 We also show the results of fitting. (They are shown as the curve in
 each figure.) }
\label{fig:explong2}
\end{figure}

\vspace{5mm}
\begin{center}
{\large Table 1}
\vskip 2mm
\begin{tabular}{ccccc} \hline 
% {L}&$50$&$100$&$200$ \\ \hline
% {A}&$(12.03\pm 3.41)\times 10^{-2}$&$(7.62\pm 1.83)\times 10^{-2}$
% &$(4.09\pm 0.76)\times 10^{-2}$ \\ \hline
% {$\sigma_{c}$}&$(4.57\pm 10.98)\times 10^{-2}$&$(3.70\pm 8.88)\times
% 10^{-2}$&$(1.23\pm 6.59)\times 10^{-2}$ \\
% {$\eta$}&$(1.416\pm 0.057)$&$(1.411\pm 0.045)$&$(1.386\pm 0.031)$ \\ \hline
% {$\chi^{2}$/freedom}&$0.493$&$0.479$&$0.754$& \\ 
{$L$}&{$A(\times 10^{-2})$}&{$\sigma_{c}(\times 10^{-2})$}&{$\eta$}
 &$\frac{\chi^{2}}{\mbox{\small freedom}}$ \vspace{1mm} \\ \hline
 $50$ & $12.03\pm 3.41$ & $4.57\pm 10.98$ & $1.416\pm 0.057$
 & $0.493$ \\  
 $100$ & $7.62\pm 1.83$ & $3.70\pm 8.88$ & $1.411\pm 0.045$
 & $0.479$ \\  
 $200$ & $4.09\pm 0.76$ & $1.23\pm 6.59$ & $1.386\pm 0.031$
 & $0.754$ \\ \hline
 \end{tabular}
\end{center}
\vskip 5mm

 The coefficient $A$ becomes smaller as we make the system size larger.
 This means that the states become more localized in larger system.
  $\sigma_{c}$ also approaches zero as the system
 size is going to infinity.
Energy eigenvalues of states
 around second peak approach zero in larger system. 
 From these facts, we expect that
 all states at $E>0$ 
 localize and there exist isolated extended states only at $E=0$
 in infinite size system with nonzero $\sigma$.
 This is the same situation as in the case of white-noise disorder. 

From the above calculations,
we can obtain the relation between the correlation length of the random 
mass $\lambda$ and the localization length $\xi$ as follows,
 \begin{equation}
 \label{eq:spincorrelation}
 \xi = C \lambda^{\tau},
 \end{equation}
 from Eqs.(\ref{eq:lambdarel}) and (\ref{eq:sigmarel}). 
The value of $\tau$ is estimated as around $0.8$.
 
In the random bond XY model, the localization length $\xi$
is related with the correlation length of spins.
On the other hand, the impurity density and/or
aperiodicity of the positions of impurities control the 
parameter $\lambda$.
Therefore we hope that the result (\ref{eq:spincorrelation})
can be observed at least qualitatively in the random spin chains.
 
%%%%%%%%%%%%%%%%%%%%%%%%%%%%%%%%%%%%%%%%%%%%%%%%%%%%%%%%%%%%%%%%%%%%%%%%

\section{Conclusion}

In this paper we studied the random-mass Dirac fermion in one dimension
by using the TMM and IVPM.
In the first half, we explained and verified the validity of the 
IVPM by comparing the obtained results with the analytical calculations.
In the second half, we investigated the effects of the 
nonlocal correlations of the random mass by using Gaussian distribution
for distances between jumps of $m(x)$.
We obtained relation between the correlation length and localization
length.
In the subsequent papers\cite{TI,TI2}, we shall study the Dirac fermion
with long-range correlated random mass.
%%%%%%%%%%%%%%%%%%%%%%%%%%%%%%%%%%%%%%%%%%%%%%%%%%%%%%%%%%%%%%%%%%%%%%%%

%\end{multicols}

%%%%%%%%%%%%%%%%%%%%%%%%%%%%%%%%%%%%%%%%%%%%%%%%%%%%%%%%%%%%%%%%%%%%%

\end{document}